\documentclass{ws-ijmpe}
 \usepackage{url}
\usepackage[super,compress]{cite}
 \usepackage{subfig}
\usepackage{float}

\newcommand{\eeq}{\end{eqnarray}}

\def\bk{{\mbox{\boldmath$k$}}}

\def\br{{\mbox{\boldmath$r$}}}
\def\b0{{\mbox{\boldmath$0$}}}

\def\bk{{\mbox{\boldmath$k$}}}

\def\br{{\mbox{\boldmath$r$}}}
\def\b0{{\mbox{\boldmath$0$}}}
\def \b #1{ {\bf #1}}
\newcommand{\bea}{\begin{eqnarray}}
\newcommand{\eea}{\end{eqnarray}}

\def \b #1{ {\bf #1}}

\def\lsim{\mathrel{\rlap{\lower4pt\hbox{\hskip1pt$\sim$}}
    \raise1pt\hbox{$<$}}}         
\def\gsim{\mathrel{\rlap{\lower4pt\hbox{\hskip1pt$\sim$}}
    \raise1pt\hbox{$>$}}}         

\def\s0{\sigma_0(s)}

\def\Vec#1{\mbox{\boldmath $#1$}}

\newcommand{\be}{\begin{equation}}
\newcommand{\ee}{\nonumber\end{equation}}

\def\s{\mbox{\boldmath $s$}}

\def\tt{\mbox{\boldmath $\tau$}}

\begin{document}

\markboth{M. Alvioli et al.}{Universality of nucleon-nucleon short-range correlations and
nucleon momentum distributions}

\catchline{}{}{}{}{}
\vskip -1cm
\title{Universality of nucleon-nucleon short-range correlations and
nucleon momentum distributions}
\author{Massimiliano Alvioli}
\address{ CNR-IRPI, Istituto di Ricerca per la Protezione Idrogeologica,  Via Madonna Alta 126, I-06128, Perugia, Italy}
\author{\footnotesize Claudio Ciofi degli Atti\footnote{Contact person: ciofi@pg.infn.it}}
\address{Istituto Nazionale di Fisica Nucleare, Sezione di Perugia,
c/o Department of Physics, University of Perugia,
   Via A. Pascoli, I-06123, Perugia, Italy}
\author{\footnotesize Leonid P. Kaptari}
\address{Bogoliubov Lab. Theor. Phys., JINR, 141980, Dubna,
 Russia}
\author{\footnotesize Chiara Benedetta   Mezzetti}
\address{Department of Chemistry and Industrial Chemistry, University of Pisa,
Via Risorgimento 35, I-56126, Italy, and\\
Consorzio Interuniversitario Nazionale per la Scienza e Tecnologia dei Materiali,
 via G. Giusti 9, Pisa, I-50121, Italy, and
 Istituto Nazionale di Fisica Nucleare, Sezione di Perugia,
  Via A. Pascoli, I-06123, Perugia, Italy}

\author{\footnotesize Hiko Morita}
\address{Sapporo Gakuin University, Bunkyo-dai 11, Ebetsu 069-8555,
  Hokkaido, Japan}
\maketitle
\begin{history}
\end{history}
\vskip-0.9cm
\begin{abstract}
By analyzing recent microscopic  many-body calculations of few-nucleon systems and complex nuclei  performed by different groups in
terms of realistic nucleon-nucleon (NN) interactions,
it is shown that NN
short-range correlations (SRCs) have a
universal character, in that the correlation hole
that they produce in nuclei
appears to be almost A-independent and similar to the correlation hole in the  deuteron.  The correlation hole creates
 high-momentum
components, missing in a mean-field (MF) description and  exhibiting several scaling properties and a peculiar spin-isospin structure.
In particular,
the momentum distribution of a pair of nucleons in spin-isospin state $(ST)=(10)$, depending upon the pair relative ($k_{rel}$) and  center-of-mass
(c.m.) ($K_{c.m.}$) momenta,  as well as upon the angle $\Theta$ between them, exhibits a remarkable property: in the region
$k_{rel}\gtrsim 2\,fm^{-1}$ and $K_{c.m.}\lesssim 1\,fm^{-1} $, the relative and c.m. motions are decoupled and the
two-nucleon momentum distribution factorizes into the deuteron momentum distribution and an A-dependent momentum distribution
describing the c.m. motion of the pair in the medium. The impact of these and other properties of one- and two-
nucleon momentum distributions on various nuclear phenomena, on ab initio calculations in terms of low-momentum interactions, as well as
 on ongoing experimental investigations of
SRCs,  are  briefly commented.\\

\noindent{\it Keywords}: Many-body approaches; NN interactions; Short Range Correlations;
 Momentum Distributions
\end{abstract}
\maketitle

\section{Introduction}
\label{sec:1}

It is  well known that many low-energy properties of nuclei can be successfully  explained in
terms of the independent  motion of nucleons in a MF created by their mutual
interaction (see, e.g. \cite{Bohr_Mottelson}). Recently, however,  it  became possible to
investigate nuclear structure at high values of energy and momentum transfers, probing
inter-nucleon distances of the order of the nucleon radius ($\simeq 1 fm$) (see e.g.
\cite{Arrington:2012} and references therein quoted). This would make it possible to
answer longstanding questions concerning the structure of nuclei at short distances, e.g.:
\begin{enumerate}
\item{\emph{what are the quantitative limits of validity of the MF picture of nuclei?}}
\item {\emph{Does the strong short-range repulsion characterizing  modern NN
 interactions \cite{RSC,Paris_NN,AV18,Pudliner:1997ck,AV6}
 manifest itself in strong NN SRCs in the
nuclear medium, i.e. strong deviations from the independent particle motion (IPM) at short
inter-nucleon distances? Are
SRCs limited to two-nucleon correlations, reminiscent of the ones occurring in
 the deuteron, or  many-nucleon SRCs  should also be considered?}}
\item{\emph{Do nucleon and meson  remain the
dominant effective degrees of freedom (d.o.f.) in the short-region domain of nuclei, or
quark and gluon d.o.f.   have to  be taken explicitly into account?}}
\item{\emph{Do the details of the short-range structure of nuclei affect unconventional
nuclear processes like, e.g., the structure of cold hadronic matter at high densities  or
 high-energy processes like nucleon-nucleus and nucleus-nucleus scattering at relativistic energies?}}
\end{enumerate}
   Unveiling the details of the  short-range structure of nuclei is a fundamental task of
 nuclear physics. As a matter of fact, it should be kept in mind  that the strong
 repulsive core in the NN potential,
resulting from the analysis of NN elastic scattering data, is introduced by means of
various form factors that leave a certain degree of arbitrariness, leading
to different short-range behaviors of various  NN interaction models.
   Moreover,
  elastic on-shell NN scattering cannot in principle
determine the details of the NN interaction in  medium, because two  nucleons that
experience
 interaction with surrounding
partners, are off-the-energy shell. As a result, a family of different phase-equivalent
 potentials
 can be derived (see, e.g. \cite{Machleidt,Epelbaum,Polyzou}) that may produce  different behaviors
 of the nuclear wave function at short distances (see e.g. \cite{Vary}).
  It should also be stressed that  recent {\it ab initio}
  many-body approaches (e.g.
  the Unitary Correlation Operator
 \cite{Roth:2010bm} or   the No-Core Shell Model  \cite{Nocore_SM} ones) that successfully
describe many low-energy properties of nuclei,
  are based upon various
 renormalization group (RG) methods (see e.g. Ref. \cite{Suzuki_ren_group,Schwenk_ren_group,Schwenk_2})
producing  phase equivalent soft NN interactions  allowing one to readily
diagonalize
  the many-nucleon Hamiltonian that would  be extremely difficult  to diagonalize by using  the
 original bare interaction. In these approaches, if high-momentum  properties have to be evaluated it is necessary
to evolve high-momentum operators within a low-momentum theory, which is no easy task, though important progress
 is being done recently \cite{Anderson,Bogner,Furnstahl:recent}. It is not the aim of this review to discuss these
  approaches, as well  modern many-body theories (for a recent review see Ref. \cite{Weise:2013})  based upon
  effective interactions derived from
 chiral perturbation theory (see e.g. \cite{Machleidt,Epelbaum}), where  short-range dynamics is described in
 terms of contact interactions amongst nucleons. In the present
 report we focus on the  effects produced by the free short-range NN interaction on various nuclear properties and
 phenomena, i.e. we focus on SRCs,  whose
 theoretical and experimental investigations are  ultimately aimed at  providing
information on  the details of in-medium short range NN dynamics.

 The importance of studying  SRCs was stressed more than fifty years ago (see e.g. \cite{Jastrow:1955,Gottfried:1963})
  but  it was only recently that, thanks to
   the enormous  progress made   by  many-body theories and experimental techniques, the theoretical and experimental studies of  SRCs
  were placed
   on  robust grounds.

 This report is mainly addressed  at providing a critical overview  of recent theoretical calculations
   demonstrating a universal character
 of SRCs, in that: (i) in coordinate space they  produce   in the two-nucleon density
 at small relative distances a
correlation hole (a region   not accessible
 to nucleons),  exhibiting, apart from normalization factors,  very mild dependence upon  the atomic weight  $A$ and
 essentially  resembling the  correlation hole   in the
 deuteron; (ii) the correlation hole, in turn,   generates
 in  the momentum distributions high-momentum components,
 missing in  MF  momentum distributions, and also
  exhibiting, to a large extent, independence upon A  and several interesting scaling properties.
  Our report is organized as follows: in Section \ref{sec:2} a  review is presented of modern
  many-body approaches to the calculation of nuclear properties in terms of realistic NN interactions and their
  prediction about the short-range structure of nuclei; Section \ref{sec:3} shows  how the action of SRCs
  affects the number of NN pairs in a given spin (S) and isospin (T) state $(ST)$; an exhaustive illustration
of the properties and spin-isospin structure of the one-body momentum distributions,
related to to the spin-isospin  structure of SRCs is presented in Section \ref{sec:4};
calculations of two-body momentum distributions are reviewed in Section \ref{sec:5}, and
one- and two-nucleon spectral functions are briefly discussed in Section \ref{sec:6}; the
Conclusions are presented in Section \ref{sec:7}.
\section{Ab initio solutions of the nuclear many-body problem
and theoretical predictions of SRCs in configuration space}
\label{sec:2}
\subsection{The standard model of nuclei}
\label{subsec:2.1}
A description of nuclei in terms of quark and gluon d.o.f.  implies the
  solution of non perturbative  QCD problems, a very difficult
 and yet unsolved task. However, as in  the case of various  many-body systems composed of
 particles having their own structure, many-nucleon systems could be viewed as systems
 of point-like particles interacting via
 proper effective interactions that incorporate the leading d.o.f. of the system
 that, in case of nuclei, are  the nucleon and exchanged  boson ones.
 However,
the reduction of  a field theoretical problem to a non-relativistic potential
 description  generates two-,
three-,$\dots$, A-body interactions, so that the general potential energy  operator assumes
the following form
\begin{equation}{\widehat V}({\bf x}_1,{\bf x}_2,{\bf x}_3,\,\dots,\, {\bf x}_A)= \sum_{n=2}^A
{\hat v}_n({\bf x}_1,\dots {\bf x}_n),
 \label{Eq:Interaction}
 \end{equation}
  \noindent
where  $ {\bf x}_i \equiv\{{\bf r}_i, {\bf s}_i, {\bf t}_i,  \}$  denotes the nucleon
generalized  coordinate, including spatial, spin and isospin coordinates.
 The relative weight of the various components in Eq. (\ref{Eq:Interaction})
 has been estimated many years ago
in Ref. \cite{Primakoff}, arguing that the relative strength between two- and
\emph{n}-body interactions should  obey the following qualitative relation
\bea
 (n-body \,\,potential)\simeq
 {\left( \frac{v_N}{c} \right)}^{(n-2)}\times (two-body\,\,potential),
 \label{uno}
\eea where $v_N$ denotes the average nucleon velocity in a nucleus and $c$ the velocity
of light. Taking $v_N \simeq 0.1c$, one is led to the conclusion that the two-nucleon
interaction is the dominant one. Though such a statement is qualitatively correct, it is
nowadays well established that three-nucleon potentials have to be considered in order to
explain the ground-state energy of light nuclei
\cite{Nogga:3_4_NF,Kievsky:1,Wiringa_review}, with four-nucleon interactions playing only
a minor role \cite{Nogga:3_4_NF} (for a recent review on three- and more-nucleon forces
see Ref. \cite{Schwenk_3NF}). Therefore the non-relativistic Schr\"{o}dinger equation
assumes the following form
\bea
\hspace{-0.3cm}
\left[
      \sum_i \,\frac{\hat{\bf p}_i^2}{2\,m_N}\,+\,\sum_{i<j}
      \,\hat{v}_2({\bf x}_i,{\bf x}_j)+\sum_{i<j<k}
      \hat{v}_3({\bf x}_i,{\bf x}_j,{\bf x}_k) \right]
      \,\Psi_f^A(\{{\bf x}\}_A)=
E_f^A\,\Psi_f^A(\{{\bf x}\}_A),
      \label{Schroedinger}
      \eea
 where   $\{{\bf x}\}_A \equiv
\{{\bf x}_1,{\bf x}_2,{\bf x}_3,\,\dots,\, {\bf x}_A$\} denotes the set of A generalized
coordinates (the spatial coordinates satisfying  the condition $\sum_{i=1}^A \Vec{r}_i
=0$) and  $f$ denotes the complete set of quantum numbers of state $f$. Eq.
(\ref{Schroedinger}) will be referred to as the {\it Standard Model} of nuclei and in
what follows we will be mainly interested in the ground-state wave function
$\Psi_{f=0}^A\equiv \Psi_0$. Once the interactions are fixed, Eq. (\ref{Schroedinger}) should
be solved {\it ab initio}, i.e. without any significant approximation which could mask or
distort the main features of $\Psi_n$.  In what follow we will consider modern 2N bare
 interactions having the following general form
\bea
\hat{v}_2({\bf x}_i,{\bf x}_j)\,=\,\sum_{p=1}^{m}\,v^{(p)}(r_{ij})
            \hat{\mathcal{O}}^{(p)}_{ij}\,\,\,\,\,\,\,\,\,\,\,\,\,\,\,\,\,r_{ij}\equiv
            |\Vec{r}_i
-\Vec{r}_j|, \label{2Ninteraction} \eea
 like, e.g., the AV18  \cite{AV18} ($m=18$) and
 AV8$^{\prime}$
 \cite{Pudliner:1997ck} ($m=8$) interactions,
 whose main components are:
 \bea &&{\cal O}^{(1)}_{ij}\,=\,1,\,\,\, {\cal
O}^{(2)}_{ij}\,=\,{\bm{\sigma}}_i\cdot{\bm{\sigma}}_j,\,\,\,
 {\cal O}^{(3)}_{ij}\,=\,{\bm{\tau}}_i\cdot{\bm{\tau}}_j,\,\,\,\,
{\cal
O}^{(4)}_{ij}\,=\,({\bm{\sigma}}_i\cdot{\bm{\sigma}}_j)({\bm{\tau}}_i\cdot{\bm{\tau}}_j)\nonumber\\
&&{\cal O}^{(5)}_{ij}\,=\,\hat{S}_{ij}, \,\,\,\, {\cal
O}^{(6)}_{ij}=\hat{S}_{ij}{\bm{\tau}}_i\cdot{\bm{\tau}}_j, \label{Operators} \eea with
\bea \hat{S}_{ij}=3(\hat{{\bm{r}}}_{ij}\cdot{\bm{\sigma}_i})
(\hat{{\bm{r}}}_{ij}\cdot{\bm{\sigma}_j})-{\bm{\sigma}_i}\cdot{\bm{\sigma}_j}
\label{tensor}.
\eea
As for 3N potentials,  several models have been proposed in order
to
 reproduce the binding energy of few-nucleon systems, that  are
underbound by about $0.2-0.3$ MeV per particle when only 2N interactions are used (see
e.g.\cite{UIX_3NF}). Within the  MF approximation,  $\sum_{i<j}
      \,\hat{v}_2({\bf x}_i,{\bf x}_j)+\sum_{i<j<k}
      \hat{v}_3({\bf x}_i,{\bf x}_j,{\bf x}_k) \Rightarrow \sum_i U({\bf x}_i)$, the ground-state solution of Eq.
(\ref{Schroedinger}) is an antisymmetrized  product  of single particle wave functions
$\phi_{\alpha_i}$, i.e.
\bea
\Psi_0(\{{\bf x}\}_A)\Rightarrow \Phi_0(\{{\bf x}\}_A)=\mathcal{\hat{A}}\prod_i^A \phi_{\alpha_i}({\bf x}_i)=
\Phi_{0p0h}(\{{\bf x}\}_A), \label{Mean_Field_WF}
\eea
where $\Phi_{0p0h}(\{{\bf x}\}_A)$
is a Slater determinant with zero particle, zero hole (0p-0h) excitations, i.e.
 with all states below the Fermi (F) level  occupied
and those above it empty ($\phi_{\alpha_i}=0$ if ${\alpha_i}
>{\alpha_F}$).
The general solution of Eq. (\ref{Schroedinger})  includes, on the opposite,  a huge
number of Slater determinants describing $np$-$nh$ excitations generated by SRCs
\bea \Psi_0(\{{\bf x}\}_A)=c_0 \,\phi_{0p0h}(\{{\bf x}\}_A) +c_1\,\Phi_{1p1h}(\{{\bf
x}\}_A)+ c_2\,\Phi_{2p2h}(\{{\bf x}\}_A)+\dots \label{particle_hole}. \eea {\it Ab
initio} direct solutions of Eq. (\ref{Schroedinger}) in terms of bare realistic
interactions are possible in the case of few-nucleon systems  (A=3, 4) within
several approaches
 (see e.g. Refs. \cite{Gloeckle:1995jg,Kievsky:1992um,Akaishi:1988,Schiavilla_old,Suzuki_Varga,Varga:1995dm,Suzuki_1,Trento:abinitio,Trento:Review}).
 For complex nuclei the fully {\it ab initio} solutions are still difficult to obtain, but for   $A \leq 12$
ground-state energies and excitation spectra were obtained  with the AV18 NN interaction
plus 3N potentials, by means of  the Green Function Monte Carlo (GFMC) method (see e.g.
Ref.\cite{Pieper_GFMC}); for $^{16}$O  the  Variational Monte Carlo (VMC) method has been
used \cite {Pieper}, and  for $A \geq 16$ the {\it cluster expansion}
approach  has been adopted with success
\cite{Arias_de_Saavedra:2007qg,Alvioli:2005cz}. The picture that emerges from these
calculations is a structure of the  ground-state wave function in the following form
\bea \Psi_0(\{{\bf x}\}_A)\,=\,{\hat F}(\{{\bf x}\}_A)\,\Phi_0(\{{\bf x}\}_A)
\label{corrwavefun}, \eea where
\bea \hat{F}(\{{\bf x}\}_A)=\mathcal{\hat{S}}\prod_{i<j}\hat{f}_{ij}({\bf x}_i,{\bf
x}_j)=\mathcal {\hat{S}}
      \prod_{i<j}\,\left [\sum_{n=1}^{m}\,f^{(n)}(r_{ij})
    \hat{\mathcal{O}}^{(n)}_{ij} \right ]
     \label{corroperator}
     \eea
 is a correlation operator introducing SRCs
into the MF wave functions $\Phi_0$,  $\mathcal{\hat{S}}$ is a symmetrization
operator, and $\hat{\mathcal{O}}^{(n)}_{ij}$ is the same operator appearing in Eq. (\ref{2Ninteraction}). It can be seen that the many-body wave function exhibits a rich correlation
structure, the dominant SRC effects arising from the short-range repulsion and the
intermediate tensor attraction.
\subsection{The one- and two-body densities  and SRCs}
\label{subsec:2.2}
Once the many-body wave function $\Psi_0$ is at disposal, the  relevant quantities of
interest are the $n$-body density, in particular:

 \noindent 1. the one-body non-diagonal
spin-isospin independent density:
\bea
\hspace{-0.5cm}
\rho(\Vec{r}_1,\Vec{r}^{\prime}_1)= A\int
\Psi_0^*(\Vec{r}_1,\{\Vec{r}\}_{A-1})\Psi_0(\Vec{r}^{\prime}_1,\{\Vec{r}\}_{A-1})
\prod_{i=2}^A d \Vec{r}_i;
\label{1BND}
\eea
\noindent 2. the two-body non-diagonal spin-isospin independent  density
\bea \hspace{-0.5cm} \rho(\Vec{r}_1,\Vec{r}^{\prime}_1;\Vec{r}_2,\Vec{r}^{\prime}_2)=
\frac{A(A-1)}{2}\int
\Psi_0^*(\Vec{r}_1,\Vec{r}_2,\{\Vec{r}\}_{A-2})\Psi_0(\Vec{r}^{\prime}_1,\Vec{r}^{\prime}_2,\{\Vec{r}\}_{A-2})
\prod_{i=3}^A d \Vec{r}_i; \label{2BND}
\eea
\noindent 3. the non-diagonal  spin-isospin
dependent two-body density
\bea \hspace{-0.5cm} \rho_{(ST)}^{N_1N_2}
({\bf r}_1, {\bf r}_1^{\prime};{\bf r}_2, {\bf r}_2^{\prime}) = \int\psi_{0}^{A*}( \{{\widetilde{\bf
x}}\}_A) \,\sum_{i<j}{\hat P}_{ij}^{S}\,{\hat P}_{ij}^{T}\,\widehat{\rho}_{ij}({\bf r}_1,{\bf r}_1^{\prime};{\bf r}_2,{\bf r}_2^{\prime})
 \psi_{0}^{A}( \{{\widetilde{\bf
x}^{\prime}}\}_A)d\textsl{\bf X};
 \label{2BND_ST}
\eea
where $d\,\textsl{\bf X}\equiv
\prod\displaylimits_{i=1}^A d\widetilde{\Vec{\bf x}}_i
d\widetilde{\Vec{\bf x}}_i^{\prime}$
and the
non-diagonal two-body density operator is
\bea
\hspace{-0.5cm}
\widehat{{\rho}}_{ij}
({\bf r}_1,{\bf r}_1^{\prime};{\bf r}_2,{\bf r}_2^{\prime})
=\,\delta(\widetilde{{\bf r}}_i- {\bf r}_1)
\delta(\widetilde{{\bf r}}_j- {\bf r}_2)\delta(\widetilde{{\bf r}_i^{\prime}}-
{{\bf r}}_1^{\prime})
\delta(\widetilde{{\bf r}}_j^{\prime}- {\bf r}_2^{\prime})
\prod\displaylimits_{k\neq \{i,j\}}^A \delta(\widetilde{{\bf r}}_k - \widetilde{{\bf r}}_k^{\prime}).
\label{2BNDDOperator} \eea
 Here $N_1$ and $N_2$ denote the two nucleons in state $(ST)$ and ${\hat P}_{ij}^{S(T)}$
 is a projection operator in the state with spin (isospin) S(T). The one-body diagonal $\rho(\Vec{r}_1)$, two-body
diagonal $\rho(\Vec{r}_1,\Vec{r}_2)$,  half-diagonal
$\rho(\Vec{r}_1,\Vec{r}_2;\Vec{r}^{\prime}_1)$ and
$\rho_{(ST)}^{N_1N_2}({\bf r}_1,\Vec{r}_2;\Vec{r}^{\prime}_1)$ densities  can  easily be obtained from Eqs. (\ref{2BND}),
(\ref{2BND_ST}) and (\ref{2BNDDOperator}), by inserting proper $\delta$-functions into
the integrals  and properly generalizing  the operator (\ref{2BNDDOperator}) (see
Ref.\cite{Alvioli:2013}).

Let us consider the diagonal two-body density

 \bea
\rho(\Vec{r}_1,\Vec{r}_2)= \sum_{ST}\rho_{(ST)}^{N_1N_2}({\bf r}_1, {\bf r}_2)=\rho({\bf r}_{rel}, {\bf R}_{c.m.}),
\label{2BND_NOST}
\eea
where the relative (\emph{rel})  and center-of-mass (\emph{\emph{c.m.}})  coordinates are
\bea \Vec{r}_{rel}=\Vec{r}_1 -\Vec{r}_2 \equiv \Vec{r}\,\,\,\,\,\,\,\,\,\,\,\,\,\,
 \Vec{R}_{c.m.}=\frac{\Vec{r}_1+\Vec{r}_2}{2} \equiv \Vec{R},
\label{Rel_CM_coordinates}
\eea
 and the following relation holds

\bea \int \rho(\Vec{r}_1,\Vec{r}_2)d\,{\bf r}_1\,d\,{\bf r}_2
=\sum_{ST}\,\int\rho_{(ST)}^{N_1N_2}({\bf r}_1, {\bf r}_2) d\,{\bf r}_1\,d\,{\bf
r}_2=\sum_{ST}N^{N_1N_2}_{(ST)}=\frac{A(A-1)}{2},
 \label{N_ST} \eea
 where  $N_{(ST)}^{N_1N_2}$ is the
number of NN pairs in state $(ST)$. The relative and \emph{c.m.} two-nucleon densities
can then be defined as follows
\bea \rho_{rel}(\Vec{r})= \int\rho(\Vec{r},\Vec{R})\,d \Vec{R}
\label{ro_rel}\,\,\,\,\,\,\, \rho_{c.m.}(\Vec{R})= \int\rho(\Vec{r},\Vec{R})d\,
\Vec{r}. \label{Ro_rel_CM} \eea
 The
knowledge of the one- and two-nucleon densities allows one to calculate various nuclear properties, e.g.
the ground-state
 energy and the momentum distributions. The various spin-isospin dependent and independent densities
 have been calculated by
 various authors in terms of {\it ab initio} or, anyway, realistic solutions of Eq. (\ref{Schroedinger})
 with bare NN realistic interactions. These, which  will be discussed  in the next Section, provide a
 very clear definition  of SRCs and their effects on NN densities in nuclei.
\subsection{The correlation hole  in
few-nucleon systems and complex nuclei}
\label{subsec:2.3}
\emph{Ab initio} calculations with bare realistic
interactions show that, apart from an obvious  normalization factor
counting the different
\begin{figure}[tbph!]
\begin{center}
\includegraphics[height=6cm]{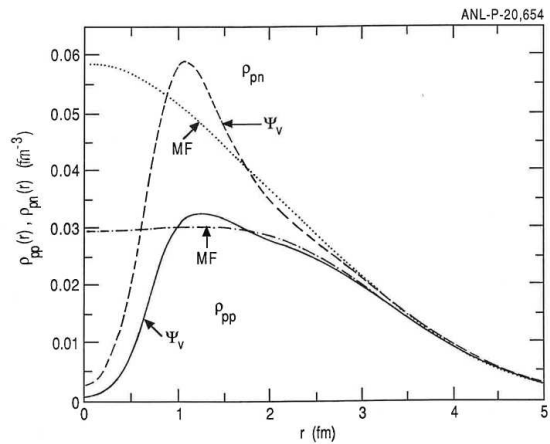}
\vspace{-0.3cm}
\caption{ The two-body density distribution of $pn$  and $pp$ pairs in $^{16}$O
corresponding to mean-field (MF) and correlated ($\Psi_V$) wave functions obtained within
the Variational Monte Carlo  approach with  AV14 NN interaction plus 3N forces (Figure reprinted from. \cite{Pieper}.
Copyright (1992) by the American Physical Society).} \label{Fig1}
\end{center}
\end{figure}
\begin{figure}[tbph!]
\hspace{-0.5cm}%
\begin{minipage}[c]{.4\textwidth}
\centering
\includegraphics[scale=0.33]{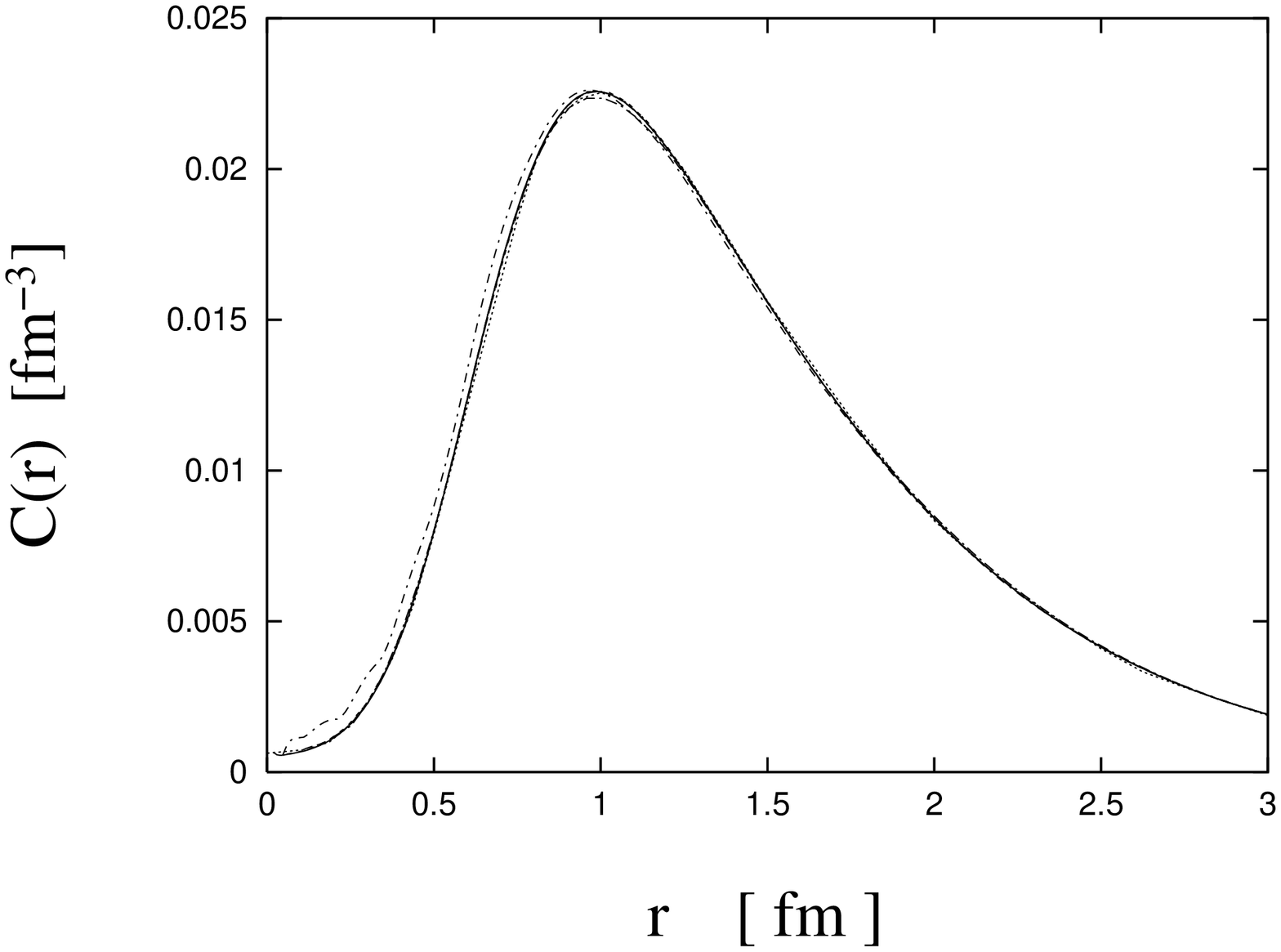}
\end{minipage}%
\hspace{1.5cm}%
\begin{minipage}[c]{.4\textwidth}
\centering
\vspace{-0.45cm}
\includegraphics[scale=1.1]{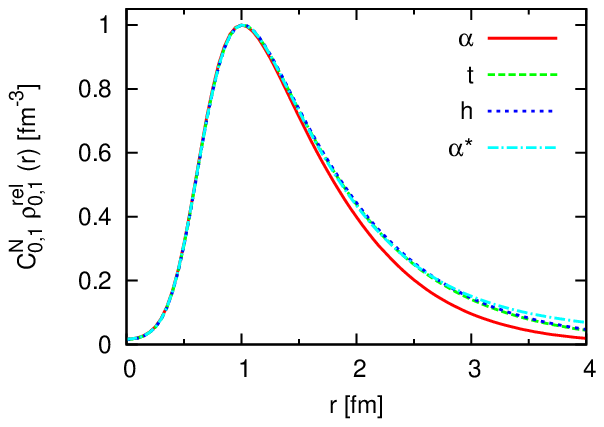}
\end{minipage}
\vspace{-0.5cm} \caption{({\bf Left}): the relative two-nucleon density in $^4$He (Eq.
(\ref{Ro_rel_CM}) with $\rho_{rel}(r) \equiv C(r)$) calculated within six different
\emph{ab initio} many-body theories using the AV18 interaction yielding practically
undistinguishable results.
 (Figure reprinted from. \cite{Nogga_Benchmark}.
Copyright (1992) by the American Physical Society). ({\bf Right}): the relative
two-nucleon density  (normalized at $r \simeq 1 \,fm$ ) in $^2$H, $^3$H, $^4$He and
$^4$He$^*$ for  NN pairs  in relative S=0 and T=1 state.
 {\it Ab initio} calculations within the method of Ref. \cite{Suzuki_1} and  AV$8^{\prime}$ interaction \cite{Pudliner:1997ck}.
 (Figure reprinted from. \cite{Feldmeier:2011qy}.
Copyright (2011) by the American Physical Society).}
\label{Fig2}
\end{figure}
number of pairs in different nuclei, the relative two-body density $\rho_{rel}(r)$
and its spin-isospin components $\rho_{ST}^{N_1N_2}(r)$
exhibits at $ r\lesssim \,
1.5 \,fm$  a sharp damping with respect to the
analogous MF density.
This is exactly the
\emph{correlation hole} previously mentioned; it is
 illustrated in Fig. \ref{Fig1} for the nucleus  of $^{16}O$. The correlation hole
is generated  by the
cooperation of the short-range repulsion  and the
\begin{figure*}[!htp]
  \centerline{\hspace*{-0.2cm}
    \includegraphics[width=5.0cm,height=4.5cm]{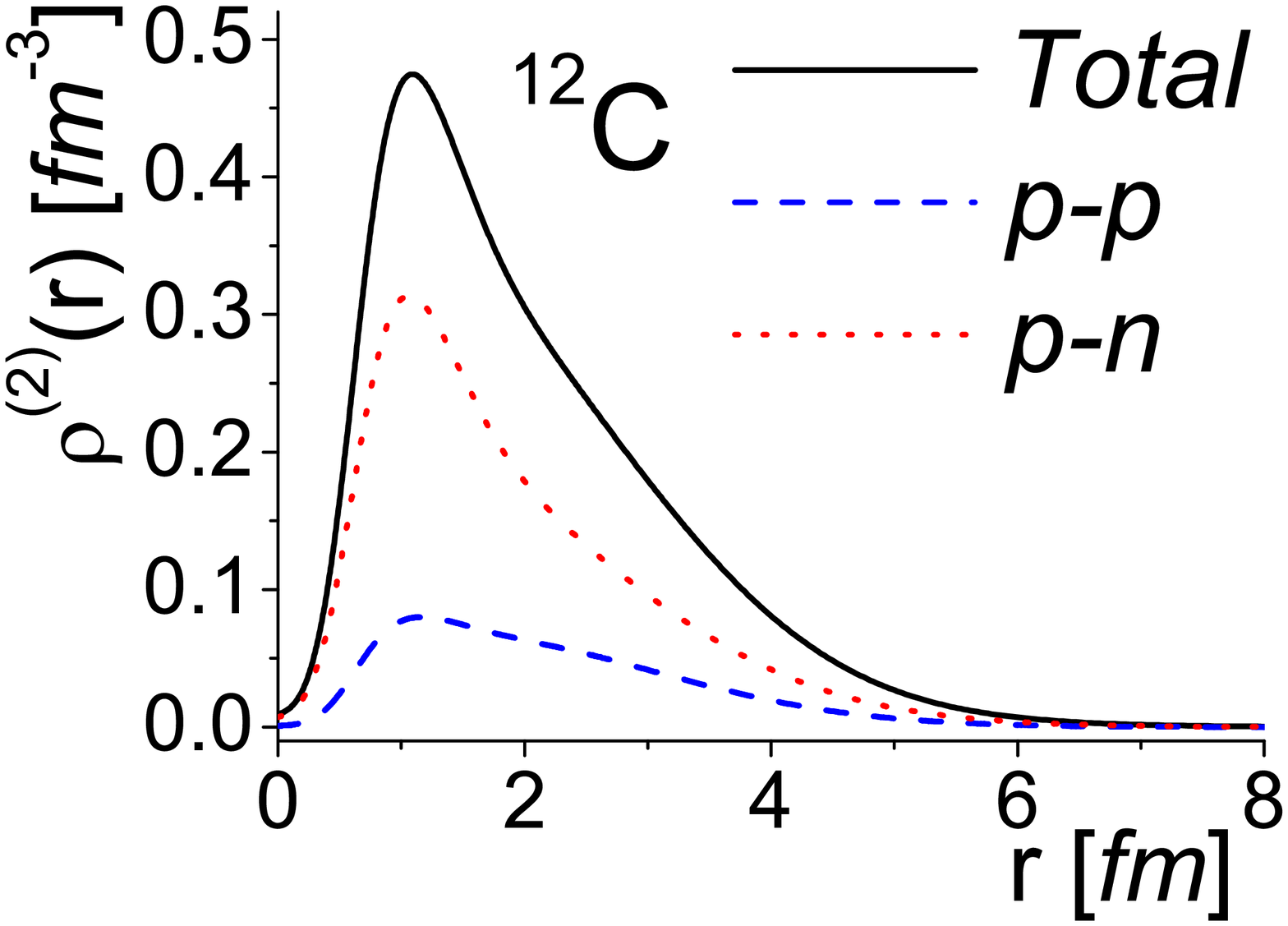}\hspace*{-0.7cm}
    \includegraphics[width=5.0cm,height=4.5cm]{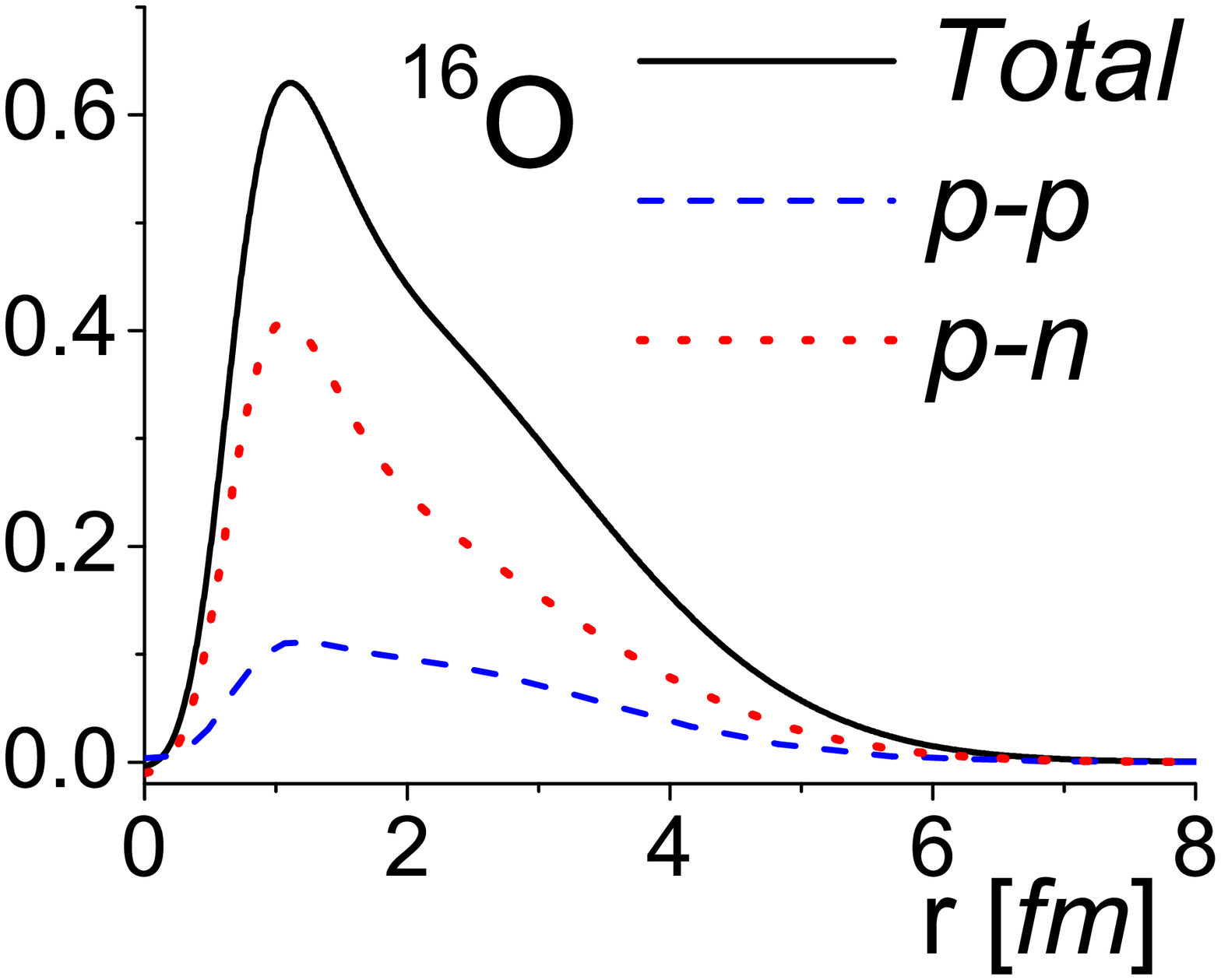}\hspace*{-0.7cm}
    \includegraphics[width=5.0cm,height=4.5cm]{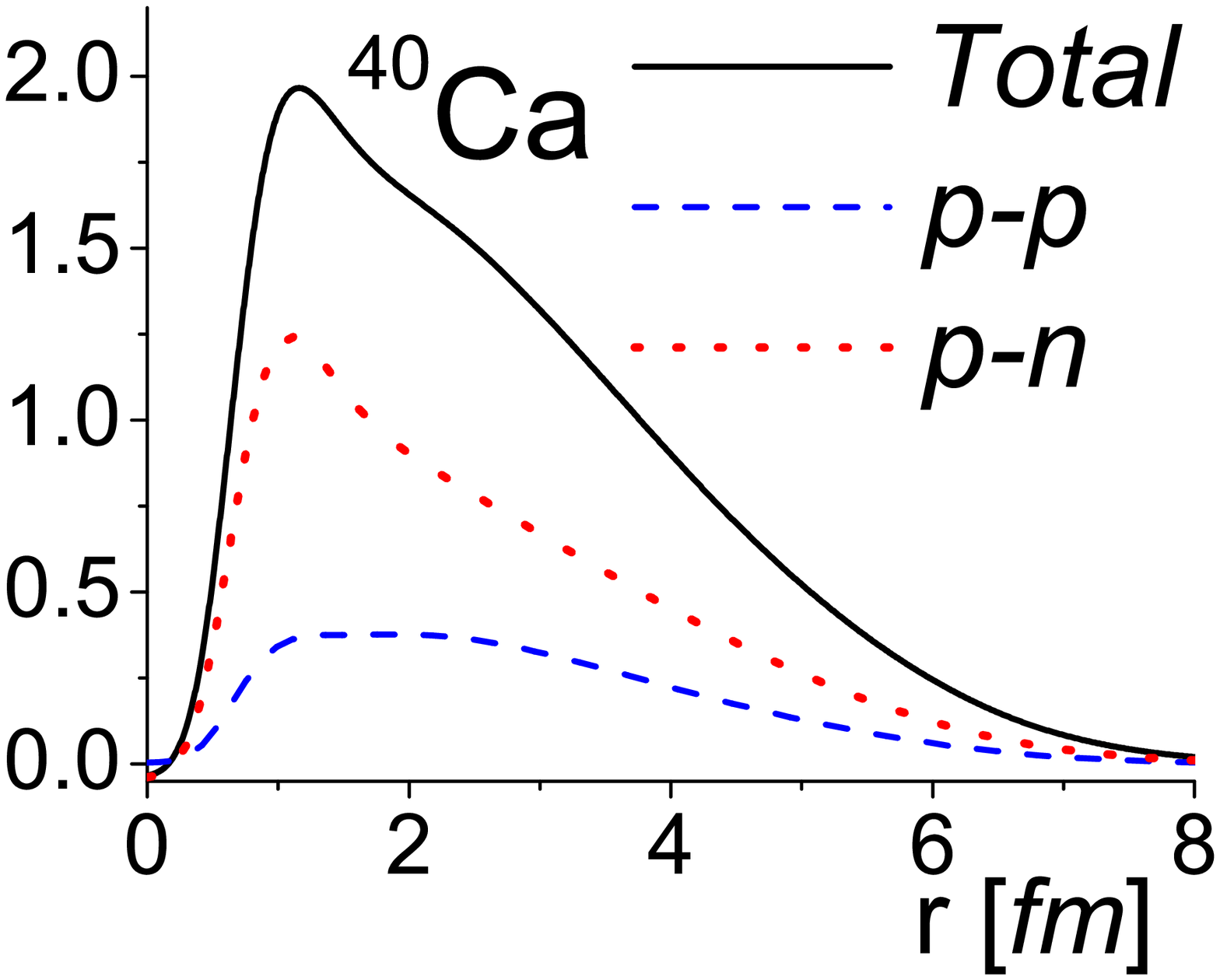}}
    \vspace{-0.3cm}
    \caption{The  two-nucleon density (Eq. (\ref{Ro_rel_CM}) with $\rho_{rel}(r)\equiv \rho^{(2)}(r)$) in  $^{12}$C, $^{16}$O, and $^{40}$Ca.
    The separate contributions of $pp$ and $nn$ densities  are also shown. The total density (full line) is given by
    $\rho^{(2)}(r)=\rho_{pn}^{(2)}(r)\,+\,2\,\rho_{pp}^{(2)}(r)$ because $\rho_{pp}^{(2)}(r)=\rho_{nn}^{(2)}(r)$. Ground-state
    wave functions from the number-conserving linked-cluster expansion calculation of Ref. \cite{Alvioli:2005cz},
    AV8$^{\prime}$ interaction
    \cite{Pudliner:1997ck}. (After Ref. \cite{Alvioli_1}).}
    \label{Fig3}
\end{figure*}
\begin{figure}
 \begin{center}
\begin{minipage}[c]{.40\textwidth}
\centering
\vspace{-0.5cm}
\hspace{-2.5cm}
 \includegraphics[scale=0.26]{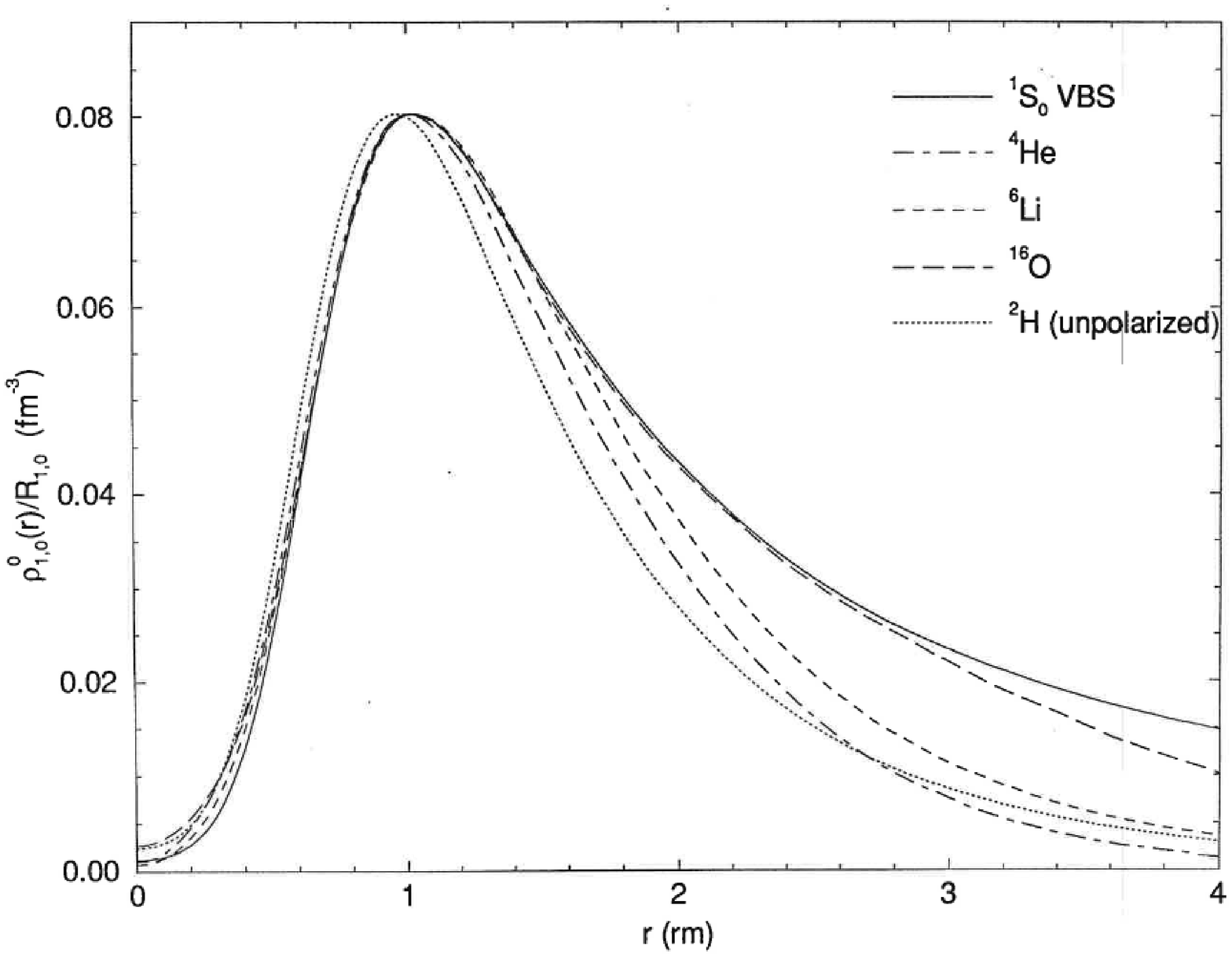}
\end{minipage}%
\hspace{-0.5cm}%
\begin{minipage}[c]{.40\textwidth}
\centering
  \includegraphics[scale=0.25] {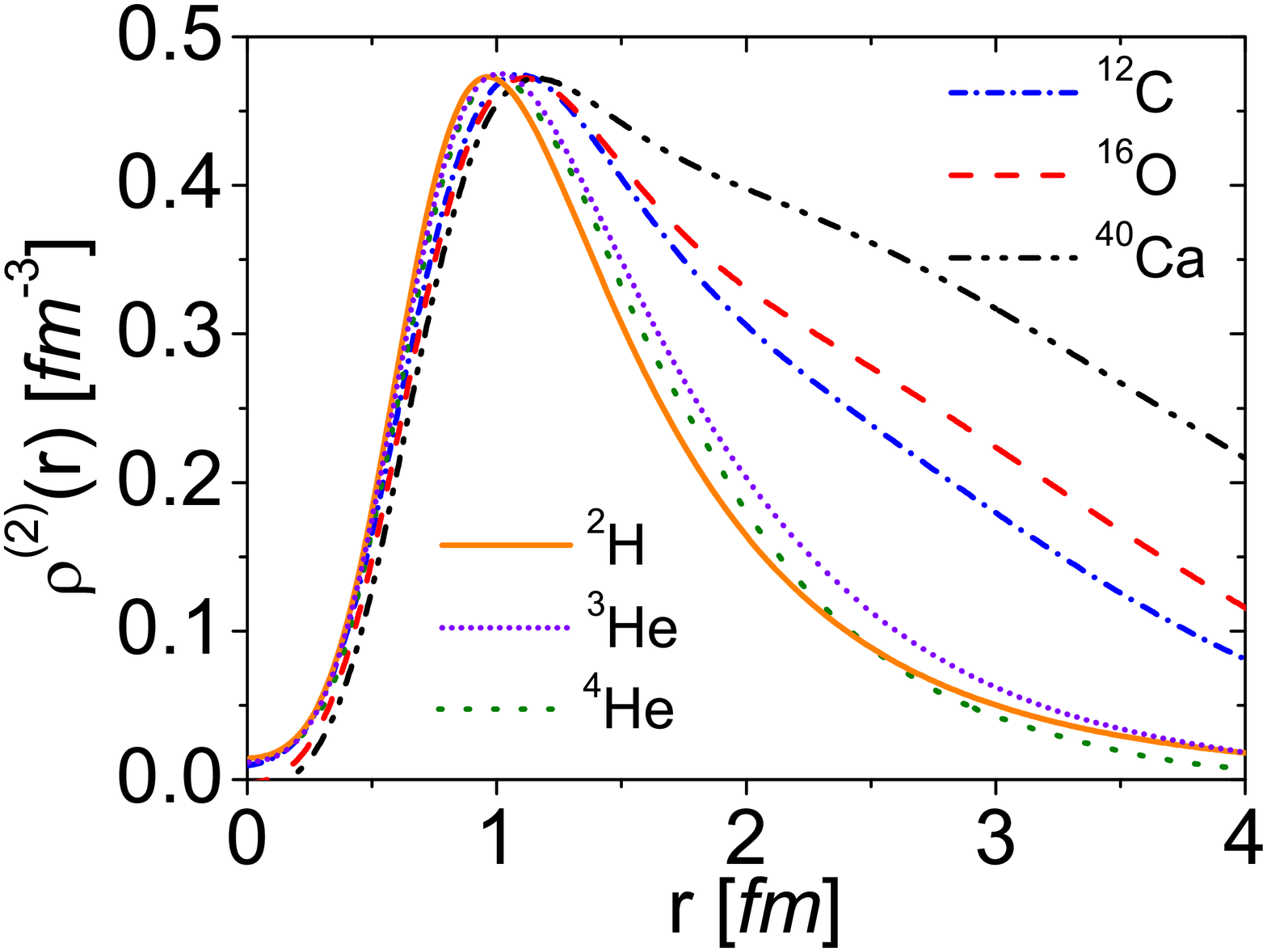}
\end{minipage}%
\end{center}
\vspace{-0.3cm}
\caption{ ({\bf Left}): the  relative two-nucleon  density (Eq. (\ref{Ro_rel_CM}) normalized at $r \simeq 1\,fm$) in $^2$H, $^4$He, $^{6}$Li and  $^{16}$O
obtained in Ref. \cite{Forest:1996kp} within the VMC method and AV18 interaction. (Figure reprinted from. \cite{Forest:1996kp}.
Copyright (1996) by the American Physical Society). ({\bf Right}):
 the two-nucleon density (Eq. (\ref{Ro_rel_CM})  with $\rho_{rel}(r)\equiv \rho^{(2)}(r)$))
 obtained with {\it ab initio} wave functions for $^3$He and $^4He$ and  within  the number-conserving linked-cluster expansion
 of Ref. \cite{Alvioli:2005cz} and the AV$8^{\prime}$ interaction \cite{Pudliner:1997ck} for $^{12}$C, $^{16}$O and $^{40}$Ca .}
\label{Fig4}
\vspace{-0.5cm}
\end{figure}
 intermediate-range tensor attraction of the NN interaction, with the tensor force  governing the overshooting  at $ r\simeq
1.0 \,fm$ in the \emph{np} distribution. Figs. \ref{Fig2}-\ref{Fig4} illustrate
the universality of the correlation hole, i.e. its
  independence
upon  $A$. These Figures also demonstrate  that different many-body approaches, ranging from
the GFMC to proper cluster expansion methods, which may give different results for the ground-state
energy,  but  predict, practically,  the same behavior of the
correlation hole. In order to be able to obtain information about this important feature
characterizing the relative NN motion in medium,  we have first of all to shift to momentum
space, expecting: (i) an increase of nucleon high-momentum components in the
ground-state wave function,  (ii) peculiar momentum configurations that are missing in a
mean-field description, and, eventually, (iii) a variation of the spin-isospin structure
of the ground-state wave function. Let us start by discussing the spin-isospin structure
of nuclei and how it is affected by SRCs.
\section{The spin-isospin structure of the nuclear ground state and SRCs}
\label{sec:3}
\subsection{The number of spin-isospin pairs in a nucleus}
\label{subsec:3.1}
\begin{figure}[tbph!]
\begin{center}
\includegraphics[height=4.5cm]{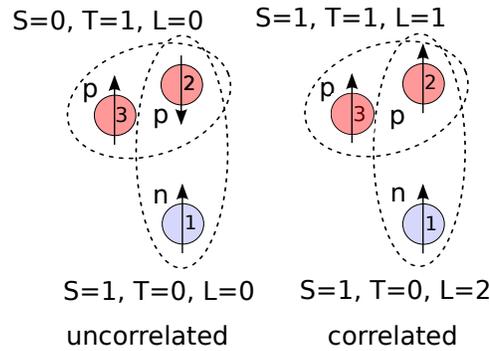}
\end{center}
\vspace{-0.5cm}
\caption{The three-body mechanism  leading to the increase of the number of pairs in $(ST)=(11)$ state (After Ref. \cite{Feldmeier:2011qy}). (Figure reprinted from
   Ref. \cite{Feldmeier:2011qy}. Copyright (2011) by
   the American Physical Society).}
\label{Fig5}
\end{figure}

The quantum numbers that characterize a two-nucleon pair in a nucleus are the relative
orbital momentum L,  the total spin S and the total isospin T. Pauli principle requires
that L + S + T={\it odd number}. In a pure shell-model picture and  $A\leq 4$  L=0, so that
(ST)=(10) and (01), whereas for $A>4$ we can have both  L even, with (ST)=(10) and  (01),
and  L  odd, with (ST)=(00) and  (11). The deviations from the shell model originating
from
SRCs, are accompanied, in $A\leq 4$
nuclei, by the creation of  $(00)$ and $(11)$ states,   and in complex nuclei  by a reduction
 of the number of $(01)$
and $(10)$ states in favor of $(11)$ and $(00)$ states.
The number of pairs in different $(ST)$ states in several nuclei  given by
\bea N_{(ST)}= \int d\Vec{r_1}\,d\Vec{r}_2
\,\rho_{(ST)}(\Vec{r}_1;\Vec{r}_2)
 \label{Enne_ST}
  \eea
  and  calculated  by
different groups,   is  reported in Table \ref{Table1}; it can be seen that:
\begin{table}[tbph!]
\begin{center}
\begin{tabular}{|c||c||c|c|c|c|}
\hline
 \multicolumn{2}{|c||}{}&\multicolumn{4}{c|}{(ST)}\\
\cline{3-6}
 \multicolumn{2}{|c||}{Nucleus}&(10)& (01)& (00)& (11)\\
\hline
$^2$H& & 1 & - & - & - \\
\hline
{$^3$He}
& IPM & 1.50 & 1.50 & - & - \\
& SRC \cite{Alvioli:2013} & 1.488 & 1.360 &0.013 & 0.139 \\
& SRC \cite{Forest:1996kp}& 1.50 & 1.350 &0.01  & 0.14  \\
& SRC \cite{Feldmeier:2011qy} & 1.489 & 1.361 & 0.011 & 0.139 \\
\hline
{$^4$He} & IPM & 3 & 3 & - & - \\
& SRC \cite{Alvioli:2013}& 2.99 & 2.57 & 0.01 & 0.43 \\
& SRC \cite{Forest:1996kp}&  3.02  & 2.5 & 0.01& 0.47\\
& SRC \cite{Feldmeier:2011qy}& 2.992 & 2.572 & 0.08 & 0.428 \\
\hline
{$^{16}$O} & IPM & 30 & 30 & 6 & 54\\
& SRC\cite{Alvioli:2013} & 29.8 & 27.5 & 6.075&  56.7 \\
& SRC \cite{Forest:1996kp}& 30.05 & 28.4 &6.05  & 55.5\\
\hline
{$^{40}$Ca} & IPM & 165& 165& 45 & 405 \\
& SRC\cite{Alvioli:2013} & 165.18 & 159.39 & 45.10 & 410.34 \\
\hline
\hline
\end{tabular}
\caption{The number of pairs $N_{(ST)}$, Eq. (\ref{Enne_ST}),
 in various
  spin-isospin states in the independent particle model (IPM) and taking into account
  SRCs within different many-body approaches (see text) with realistic interactions (AV18 and AV8'). (Table reprinted from
   Ref. \cite{Alvioli:2013}. Copyright (2013) by
   the American Physical Society)}
  \label{Table1}
  \end{center}
\end{table}
(i) SRCs do not practically affect the state $(10)$,  but appreciably
 reduce the state $(01)$, in favor  of the
 $(11)$ state; this  is ascribed  to  a three-body-like mechanism originating
 from the tensor force \cite{Forest:1996kp,Feldmeier:2011qy} illustrated in Fig. \ref{Fig5}:
tensor  correlations between particles "2" and "3" generate
a  spin flip of particle "2", that   gives rise to the state $(11)$ between particles "2"
 and "1" ; (ii) as in the case of the correlation hole, there is again
 a general agreement between the results by different groups   using different many-body approaches, namely:
the VMC   with various Argonne
interactions, in Ref.\cite{Forest:1996kp};   the
correlated Gaussian basis approach \cite{Varga:1995dm} with the
V8$^\prime$ interaction, in Ref. \cite{Feldmeier:2011qy}; the   hyperspherical
 harmonic variational method
  with
the $AV18$ interaction in Ref.\cite{Kievsky:1992um};  the ATMS  method of Ref.
\cite{Akaishi:1988} with the AV8$^\prime$ interaction and
the linked-cluster expansion of Ref.  \cite{Alvioli:2005cz} with
the AV8$^\prime$ interaction,  in Ref. \cite{Alvioli:2013}.
\section{One-body momentum distributions and SRCs}
\label{sec:4}
Let us now discuss how and to what extent SRCs affect the one-body momentum distribution,
i. e.  the Fourier transform of the non-diagonal one-body density
\bea n_A(\Vec{k}_1)=\frac{1}{A{(2\,\pi)}^3}\int
e^{-i\,\Vec{k}_1\cdot(\Vec{r}_1-\Vec{r}_1^\prime)} \rho(\Vec{r}_1,\Vec{r}_1^{\prime}) d
\Vec{r}_1 d\Vec{r}_1^{\prime} =\int n_A^{N_1N_2}({\bf k}_1, {\bf k}_2)\,
 d\Vec{k}_2
\label{Momdis} \eea
where $n^{N_1N_2}$ is the two-body momentum distribution to be discussed later on, and  $ \int
n_A(\Vec{k}_1)d\,\Vec{k}_1=1$,  which is the normalization adopted in the rest of the paper.
\subsection{General definitions and  two-nucleon SRC (2N-SRC) configurations}
\label{subsec:4.1}
SRCs   considerably increase
the high-momentum content of the one-body momentum distributions through the term
$\sum_{n=2}^{\infty}c_n\Phi_{npnh}$, in Eq. (\ref{particle_hole}), i.e. via the population
of $np$-$nh$ states with momentum much
higher
than the Fermi momentum $k_F \simeq 1.4 fm^{-1}$. SRCs, moreover,
generate peculiar wave function
\begin{figure}[tbph!]
\hspace{-1.cm}
\begin{minipage}[c]{0.95\textwidth}
\centering
\includegraphics[scale=0.45]{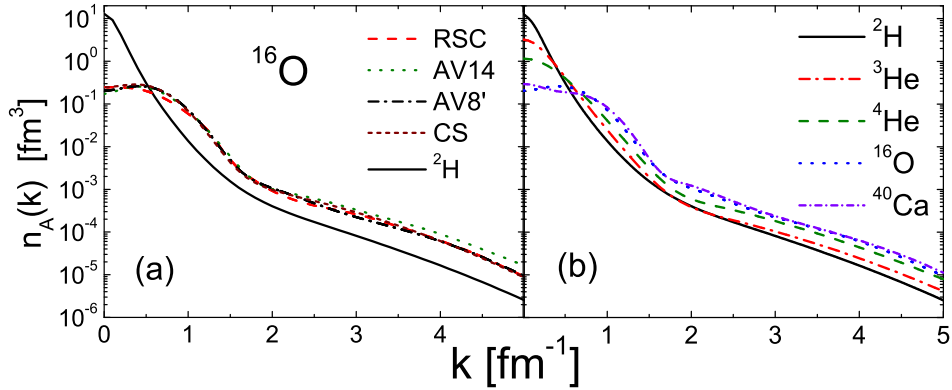}
\end{minipage}%
\vspace{-0.8cm}
\caption{ ({\bf a}): the momentum distribution in $^{16}$O calculated with different NN interactions and
theoretical approaches: RSC \cite{Zabolitzky:1978cx}; AV14 \cite{Pieper}; AV8' \cite{Alvioli:2005cz}.
The phenomenological distribution of Ref. \cite{CiofidegliAtti:1995qe} is also shown (CS) and $^2$H
denotes the deuteron momentum distribution.
  ({\bf b}): the  proton momentum distribution of   different nuclei
     calculated within different
    many-body approaches with
    equivalent NN  interactions, namely the AV18 one, in the case of
    $^2$H and $^3$He, and the AV8$^\prime$ one, in the
    case of $^4$He, $^{16}$O, and $^{40}$Ca. Hereafter the notation $|{\bf k}_1| \equiv k$ will be adopted. (Figure reprinted from. \cite{Alvioli:2013}.
Copyright (2013) by the American Physical Society).}
\label{Fig6}
\end{figure}
configurations that are missing in a MF description \cite{Frankfurt:1988nt}.  As a matter of fact, since
momentum conservation requires that
\bea \sum_{i=1}^A \Vec{k}_i=0 \label{momcons}
\eea
 a nucleon with high  momentum  $\Vec{k}_1$  in a  MF configuration
 is expected to be   balanced by the rest of the $(A-1)$ nucleons, i. e.
 \bea
 \Vec{k}_1 \simeq
-\sum_2^A \Vec{k}_i\,\,\,\,\,\,\,\,\,\,\,\,\, \Vec{k}_i \simeq \frac{\Vec{k}_1}{A-1},
\label{momcons_MF}
\eea
whereas in  a 2N-SRC configuration  one has
 \bea
 \Vec{k}_1 \simeq -\Vec{k}_2
\,\,\,\,\,\,\,\,\, \Vec{K}_{A-2}= \sum_3^A \Vec{k}_i \simeq 0. \label{momcons_2NC}
\eea
Therefore 2N-SRCs can be defined as those   configurations of a pair of
nucleons characterized by {\it high}
relative  and {\it small} \emph{c.m.}  momenta.
The quantitative meaning of such a  statement will be discussed later on.ù
\subsection{Recent calculations of the one-body momentum distribution}
\label{Subsec:4.2}
A recent systematic analysis of realistic calculations of $n_A(k)$ for A=2, 3, 4, 16, and 40 has
been presented in Ref. \cite{Alvioli:2013}.
 The results for
$^{16}$O, performed by different groups, is shown  in Fig. \ref{Fig6}({\bf a}), which is aimed at illustrating the convergence of
different  approaches that use  similar NN interactions, whereas Fig. \ref{Fig6}({\bf b}) shows that
the  high-momentum part of $n_A({\bf k}_1)$ of  different nuclei exhibits  a qualitative
universal scaling behavior. This point
will be discussed on a more quantitative level in Section \ref{subsec:4.5}.
\subsection{The probability of MF and SRC configurations}
\label{subsec:4.3}
The ground-state wave function  $\Psi_0$, solution of Eq. (\ref{Schroedinger}) describes
both  MF  and correlated-nucleon motions. The latter, in turn, includes both
long- and short-range correlations; long-range correlations (LRC) manifest themselves
mostly in open shell nuclei, and are responsible for configuration mixing
resulting in  partial occupation of states which are empty in a simple
independent particle model,
 with small effects, however,  on high-momentum components;
SRCs, on the contrary,  generate high virtual particle-hole
excitations even in closed-shell nuclei, and strongly affect
the high-momentum content of the wave function.
Therefore, assuming
 that the momentum
distributions could be extracted
from some experimental data,  we have to figure out a clear cut  way  to disentangle
the momentum content generated by the MF and LRCs from the one
 arising from SRCs. Denoting
by $\{|\psi_f^{A-1}>\}$ the complete set of  eigenfunctions of nucleus $(A-1)$
described by the same  Hamiltonian  of nucleus A,    and using the
completeness relation
 \bea
 \sum_{f=0}^{\infty}
|\Psi_f^{A-1}><\Psi_f^{A-1}|=1,
 \label{completness}
\eea
the one-nucleon momentum distribution can be written as follows \cite{CPS_PkE_momdis}
\bea
n_A(\Vec{k}_{1})= n_{gr}(\Vec{k}_{1})
+n_{ex}(\Vec{k}_{1}),
 \label{nsplit}
\eea
where
 \bea
&& \hspace{-0.7cm}(2\pi)^3 n_{gr}(\Vec{k}_{1})=\nonumber\\
&& \hspace{-0.7cm}=\sum_{f=0,\sigma_1}\Big |
\int e^{i\Vec{k}_{1}\cdot\Vec{r}_1} d\Vec{r}_1\int\chi_{\frac{1}{2}\sigma_1}^
 {\dagger}\Psi_{f=0}^{(A-1)*}
 (\{{\Vec r}_i\}_{A-1})\Psi_{0}
 ({\Vec r}_1, \{{\Vec r}_i\}_{A-1})
 \prod_{i=2}^A d\Vec{r}_i \Big |^2
 \label{ennegr}
  \eea
and
 \bea
&&\hspace{-0.7cm}(2\pi)^3 n_{ex}(\Vec{k}_{1})=\nonumber\\
 &&\hspace{-0.8cm}=\sum_{f \neq 0,\sigma_1} \Big |\int e^{i\,\Vec{k}_{1}\cdot\Vec{r}_1}
  d  \Vec{r}_1 \int \,\chi_{\frac{1}{2}\sigma_1}^
 {\dagger}\Psi_{f}^{(A-1)*}
 (\{{\Vec r}_i\}_{A-1})\Psi_{0}
 ({\Vec r}_1, \{{\Vec r}_i\}_{A-1}) \prod_{i=2}^A  d  \Vec{r}_i \Big |^2.
 \label{enneex}
 \eea
Here the sum over $f$ stands also for
 an integral over the continuum final states  that are present in Eq.
 (\ref{completness}).
 We see that the momentum distribution can be expressed through the
overlap integrals between the ground-state wave function
$\Psi_{0}$ of nucleus
 A and the wave function  $\Psi_{f}^{(A-1)}$
of the state $f$ of nucleus $(A-1)$.
  The separation of the
  momentum distributions in $n_{gr}$ and $n_{ex}$ is particularly useful
  for $A=3,\, 4$ nuclei, i.e. when the excited states of $(A-1)$ are in the continuum.
For  complex nuclei, where many discrete hole excited states are present,
  it is more convenient  to use another representation
  where the
\begin{figure}[tbph!]
\hspace{-1.cm}
\begin{minipage}[c]{0.95\textwidth}
\centering
\includegraphics[scale=0.46]{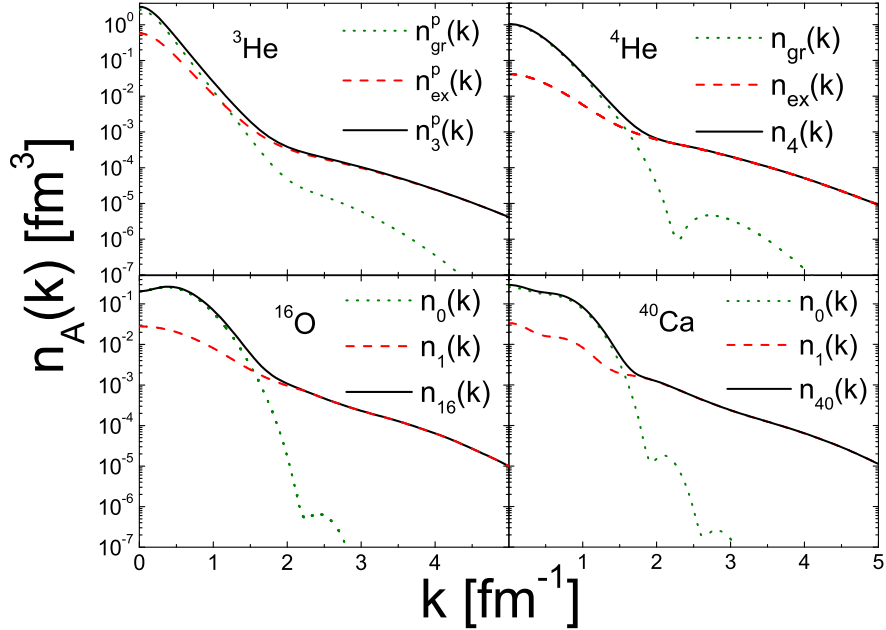}
\end{minipage}%
\vspace{-0.8cm}
\caption{The proton momentum distribution $n_A^{p}({k}_{1})\equiv n_A(k)$  and its separation into the uncorrelated
and correlated contributions, Eqs. (\ref{nsplit}-\ref{enneuno}),
 in A=3 (wave function from Ref.
\cite{Kievsky:1992um}, AV18 interaction),  A=4 (wave function from Ref. \cite{Akaishi:1988},
AV8$^\prime$ interaction), A= 16 (wave functions from Ref. \cite{Alvioli:2005cz},
AV8$^\prime$ interaction), and A=40 (wave function from Ref. \cite{Alvioli:2005cz},
AV8$^\prime$ interaction).
 The
    values of the probabilities
    $\mathcal{P}_{gr(0)}^p=4 \pi\int k^2\,dk \,n_{gr}^p(k)$ and
     $\mathcal{P}_{ex(1)}^p=4 \pi\int k^2\,dk \,n_{ex}^p$, Eq.(\ref{Probabilities}),  are listed in Table
    \ref{Table2} and the partial probabilities, Eq. (\ref{partialprob1}), in Table \ref{Table3}.
 (Figure reprinted from \cite{Alvioli:2013}.
Copyright (2013) by the American Physical Society)}
\label{Fig7}
\end{figure}
particle-hole structure of the  realistic solutions
of Eq. (\ref{Schroedinger}) is explicitly exhibited by  Eq. (\ref{particle_hole}). Within
such a representation, one has \cite{Ciofi_Liuti,Ciofi_Liuti_Simula}
\bea
n_A(\Vec{k}_{1})=
n_{0}(\Vec{k}_{1}) +n_{1}(\Vec{k}_{1}),
 \label{nsplit1}
\eea
 where
 \bea
 &&\hspace{-0.75cm}(2\pi)^{3}\,n_{0}(\Vec{k}_{1})=\nonumber\\
 &&\hspace{-0.75cm}=
  \sum_{f\leq F,\sigma_1} \Big |\int e^{i\Vec{k}_{1}\cdot\Vec{r}_1} d \Vec{r}_1
 \int \,\chi_{\frac{1}{2}\sigma_1}^
 {\dagger}\Psi_{f}^{(A-1)*}
 (\{{\Vec r}_i\}_{A-1})\Psi_{0}
 ({\Vec r}_1,\{{\Vec r}_i\}_{A-1})\prod_{i=2}^A d\Vec{r}_i \Big |^2
 \label{ennezero}
  \eea
 \bea
&&\hspace{-0.8cm}(2\pi)^3\,n_{1}(\Vec{k}_{1})=\nonumber\\
 &&\hspace{-0.8cm}=\sum_{f>F,\sigma_1}\Big |\int e^{i\Vec{k}_{1}\cdot\Vec{r}_1}
  d\Vec{r}_1 \int \,\chi_{\frac{1}{2}\sigma_1}^
 {\dagger}\psi_{f}^{(A-1)*}
 (\{{\Vec r}_i\}_{A-1})\psi_{0}
 ({\Vec r}_1, \{{\Vec r}_i\}_{A-1}) \prod_{i=2}^A d\Vec{r}_i \Big |^2.
 \label{enneuno}
  \eea
The summation over $f$ in Eq. (\ref{ennezero}) includes
all the discrete
 shell-model levels below the Fermi level in $(A-1)$
 ("hole states" of $A$),  and in
Eq. (\ref{enneuno})  it includes all the discrete and continuum states above the Fermi
level created by SRCs. In a fully uncorrelated MF approach, one has
\bea n_A(\Vec{k}_{1})= n_{0}(\Vec{k}_{1})=\sum_{\alpha \leq F} \left
|\phi_{\alpha}(\Vec{k}_{1}) \right |^2;  \qquad n_{1}(\Vec{k}_{1})= 0.
\label{nmeanfield} \eea
\begin{table}[!ht]
\begin{center}
{\renewcommand\arraystretch{1.3}
\begin{tabular}{|c||c||c|c|} \hline
\multicolumn{4}{|c|}{MEAN FIELD AND SRC PROBABILITIES} \\ \hline
$Nucleus$ &    $Potential$ &    \,\,\,$\mathcal{P}_{gr}$    & $\mathcal{P}_{ex}$  \\
\hline
{$^{3}$He \cite{Kievsky:1992um}}        & AV18 \cite{AV18}   &  0.677 & 0.323  \\ \hline
{$^{4}$He \cite{Akaishi:1988,Schiavilla_old}} & RSC \cite{RSC} AV8$^\prime$\cite{Pudliner:1997ck} & 0.8 & 0.2 \\
\hline
 $ Nucleus $  &  $ Potential$  &   $\mathcal{P}_0$   &   $\mathcal{P}_1$    \\ \hline
{$^{16}$O \cite{Alvioli:2005cz}} & V8' \cite{Pudliner:1997ck} & 0.8  & 0.2   \\ \hline
{$^{40}$Ca \cite{Alvioli:2005cz}} & V8' \cite{Pudliner:1997ck} & 0.8  & 0.2   \\
\hline
\hline
\end{tabular}}
\end{center}
\caption{The proton MF,  $\mathcal{P}_{gr(0)}^p=\int  d\,{\bf k}_1 \, n_{gr(0)}^{p}({\Vec k}_1)$,
  and SRC, $\mathcal{P}_{ex(1)}^p=\int  d\,{\bf k}_1  \, n_{ex(1)}^{p}({\Vec k}_1)$, probabilities,
  Eq. (\ref{Probabilities}), in various nuclei
  obtained from AV18 and AV8' interactions.
   (Table reprinted from. \cite{Alvioli:2013}.
Copyright (2013) by the American Physical Society).}
 \label{Table2}
\end{table}
The modulus squared   of the overlap integral represents the weight of the ground and
excited virtual states of $(A-1)$ in the ground state of $A$, so that the quantities
 \bea
 \mathcal{P}_{gr(0)}=
 \int_0^{\infty} n_{gr(0)}(\Vec{k}_{1})\,d\,\Vec{k}_1 \,\,\,\,\,\,\,
 \mathcal{P}_{ex(1)}=\int_0^{\infty}  n_{ex(1)}(\Vec{k}_{1})\,d\,\Vec{k}_1,
 \label{Probabilities}
  \eea
  with
  \bea
\mathcal{P}_{gr(0)}+\mathcal{P}_{ex(1)}=1,
\label{sumesse}
  \eea
yield, respectively,   the probability to find
a MF and a correlated nucleon in the range
$0 \leq k_1 \leq \infty$;  they can therefore be assumed as the MF and
SRC total probabilities.
It is clear that both low- and high-momentum components contribute to mean-field and correlated
momentum distributions but, as it should be expected,  $n_{gr(0)}$ ($n_{ex(1)}$) should get contribution mainly
from low (high)
momentum components. This is clearly illustrated in  Fig. \ref{Fig7},
where the proton momentum distributions
\begin{table} [!h]
\normalsize
\begin{center}
{\renewcommand\arraystretch{1.5}
\begin{tabular}{|c|c|c|c|c|c|c|c|c|c|}
\hline
 &$^2$H&\multicolumn{2}{c|}{$^3$He(p)}&\multicolumn{2}{c|}{$^4$He}&\multicolumn{2}{c|}{$^{16}$O}&\multicolumn{2}{c|}{$^{40}$Ca}\\
\hline
 $k_1^-$& $\mathcal{{P}}$&$\mathcal{{P}}_{gr}$ & $\mathcal{{P}}_{ex}$& $\mathcal{{P}}_{gr}$ & $\mathcal{{P}}_  {ex}$ & $\mathcal{{P}}_0$ & $\mathcal{{P}}_1$ & $\mathcal{{P}}_0$ & $\mathcal{{P}}_1$\\
\hline  
$0.0$ &$1.0$ & $0.7$ &  $0.3$       & $0.8$ & $0.2$    & $0.8$ & $0.2$    & $0.8$ & $0.2$\\
\hline
$0.5$ & $0.3$    &  $0.3$  &   $0.2$   & $0.5$ & $0.1$    & $0.7$ & $0.2$          & $0.7$ & $0.2$\\
\hline
$1.0$ & $0.08$  &    $0.03$ &  $0.07$  & $0.1$ & $0.1$   & $0.2$ & $0.1$    & $0.2$ & $0.1$\\
\hline
$1.5$ &   $0.06$ &  $0.005$ &  $0.04$    & $0.008$ & $0.08$   & $0.008$ & $0.1$      & $0.01$ & $0.1$ \\
\hline
$2.0$ &  $0.04$ & $0.002$ &  $0.02$  & $7\cdot10^{-4} $ & $0.06$  & $6\cdot10^{-4}$ & $0.06$  & $3\cdot10^{-4}$ & $0.07$  \\
\hline
\hline
\end{tabular}}
\caption{The values of the proton partial probability, Eq. (\ref{partialprob1}),
for $^{3}$He, $^{4}$He, $^{16}$O and $^{40}$Ca, calculated for different values of
the
momentum $k_1^-$ (in fm$^{-1}$) with  $k_1^+=\infty$.  (Table reprinted from. \cite{Alvioli:2013}.
Copyright (2013) by the American Physical Society).}
\label{Table3}
\end{center}
\end{table}
of A=3, 4, 16, and 40 nuclei are shown with the separation into the MF and correlation
contributions: it can be seen that, starting from $k\gtrsim 2\,fm^{-1}$,
 the momentum distributions are dominated by the correlated part.
The calculated values of $\mathcal{P}_{gr(0)}$ and $\mathcal{P}_{ex(1)}$ for several nuclei are listed in Table \ref{Table2}.
Assuming that   $n_0^{N_1}$ and $n_1^{N_1}$ could  be obtained from some measurable cross section, it might well be that
 only a limited range of momenta
is available experimentally, in which case it is useful to define the partial probabilities
 \begin{figure}%
\hspace{-1.cm}
\begin{minipage}[c]{0.95\textwidth}
\centering
\includegraphics[scale=0.47]{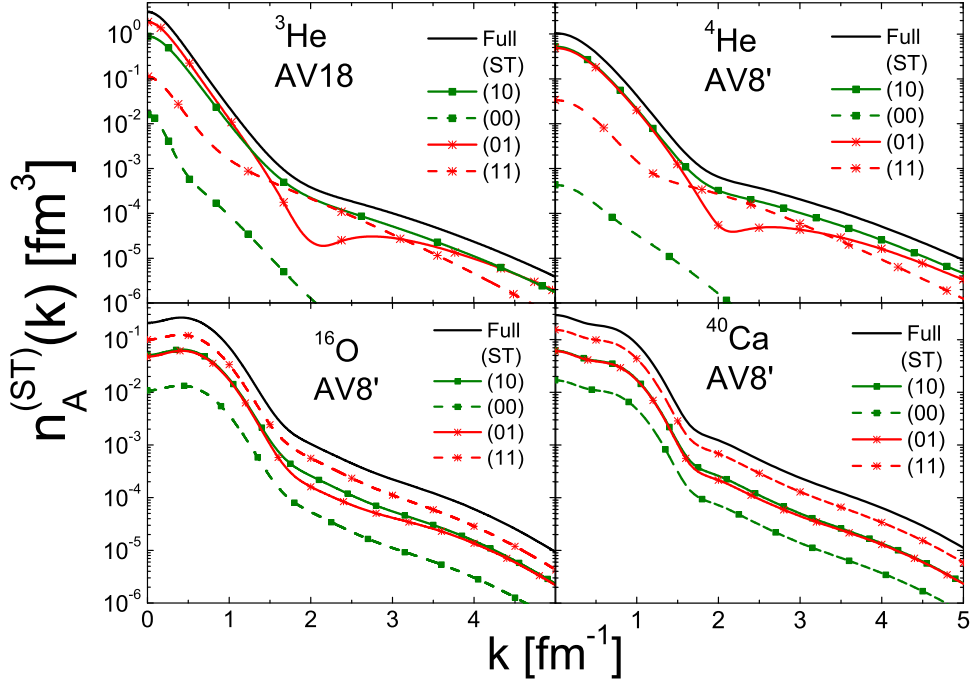}
\end{minipage}%
\vspace{-0.8cm}
\caption{The various spin-isospin contributions  to the proton distributions in $^3$He, $^4$He, $^{16}$O
 and $^{40}$Ca (Eq. (\ref{partialprob1})). Wave functions as in Fig. \ref{Fig6}.
 (Figure reprinted from. \cite{Alvioli:2013}.
Copyright (2013) by the American Physical Society)}
\label{Fig8}
\end{figure}
\bea
{\mathcal P}_{0(1)}(k^{\pm}_{1})=4\,\pi
\int_{{k^{-}_{1}}}^{{k^{+}_{1}}}
n_{0(1)}(\Vec{k}_{1}){k}_{1}^2\,d\,{k}_{1}
\label{partialprob1}
\eea
i.e.  the probability to observe a MF or a correlated nucleon with momentum
 in the range $k^{-}_{1} \leq k_1 \leq k^{+}_{1}$. The calculated values of ${\mathcal P}_{0(1)}^{N_1}(k^{\pm}_{1})$ are
 given in Table \ref{Table3}.
\subsection{The spin-isospin structure of the one-body momentum distributions}
\label{subsec:4.}
By introducing the spin-isospin dependent half-diagonal density matrix
$\rho^{N_1N_2}_{(ST)}(\Vec{r}_1,\Vec{r}_1^{\prime};\Vec{r}_2)$, the one-body
momentum distribution can be expressed in terms of its various spin-isospin components as follows
\cite{Alvioli:2013} \bea
\hspace{-0.5cm}n_A(\Vec{k}_{1})=\sum_{(ST)}n_A^{(ST)}(\Vec{k}_{1}) =\int
d\Vec{r}_1\,d\Vec{r}_1^\prime
e^{i\,\Vec{k}_{1}\cdot\left(\Vec{r}_1-\Vec{r}_1^\prime\right)}\, \sum_{(ST)}\, \int
d\Vec{r}_2 \rho^{N_1N_2}_{(ST)}(\Vec{r}_1,\Vec{r}_1^{\prime};\Vec{r}_2). \label{1BMD_ST} \eea

In Ref. \cite{Alvioli:2013} the spin-isospin dependent half-diagonal two-body density  has been calculated
for A=3, 4, 16 and 40, and the various spin-isospin contributions to $n_A(k)$
have been obtained as  shown in Fig. \ref{Fig8}. It appears  that: (i) the contribution from the $(00)$
state is negligible, both in few-nucleon systems and complex nuclei; (ii) the contribution from the $(11)$
state in $^3$He and $^4$He is small,
 both at low
and large values of $k$, but it plays a relevant role in the region
$1.5 \lesssim k \lesssim 2.5\,fm^{-1}$; (iii) in the proton distribution of $^3$He
 the $(01)$ contribution  is important  everywhere except in the
 region $1.5 \lesssim k \lesssim
3\,fm^{-1}$;
(iv) in complex nuclei, in agreement with the results shown in Table \ref{Table1},  the
 $(11)$ state (odd relative orbital momenta) plays  a dominant role, both  in the
 independent particle  model and in  many-body approaches. These observations are
useful for understanding  the material presented  in the next Section.

\subsection{The momentum distribution of nuclei {\it vs} the deuteron momentum distributions}
\label{subsec:4.5}
\begin{figure}[tbph!]
\vspace{-0.8cm}
\hspace{-1.cm}%
\begin{minipage}{.99\textwidth}
\centering
\includegraphics[scale=0.44]{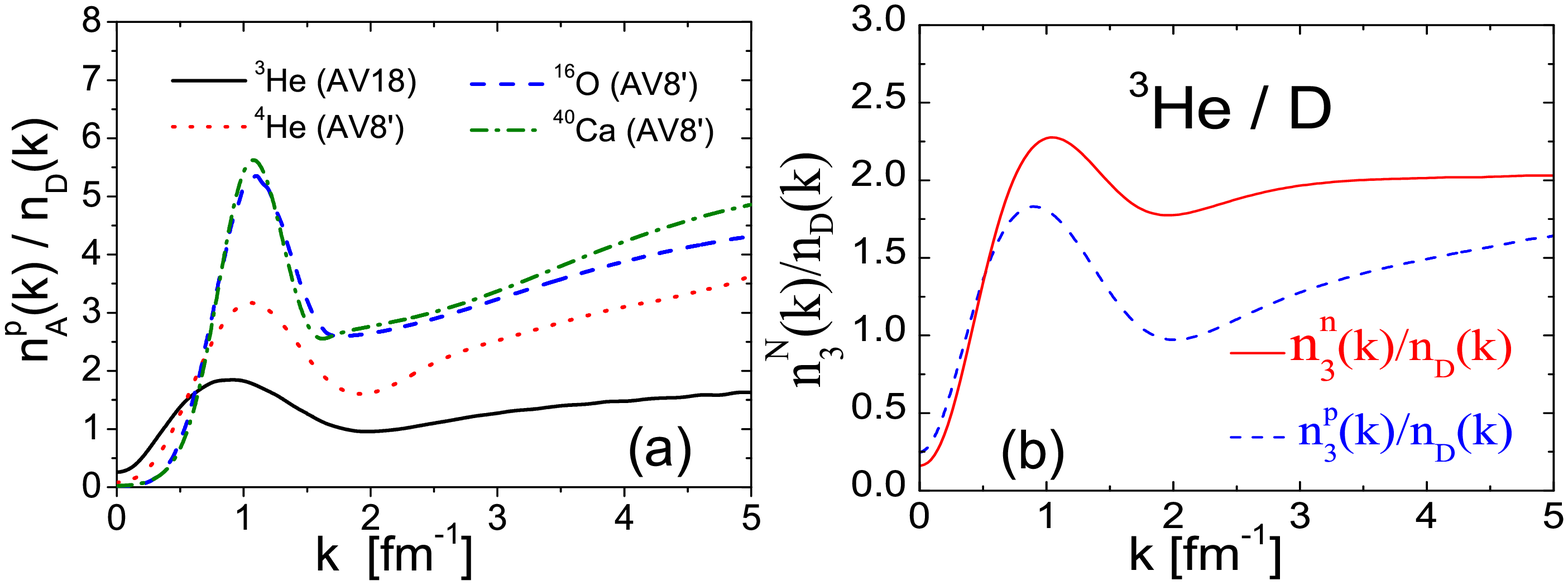}
\end{minipage}%
\vspace{-0.4cm}
\caption{({\bf a}): the ratio of the proton  momentum distribution in nucleus A, $n_A^N(k)$,  to the deuteron
momentum distribution $n_D(k)$. In  isoscalar
nuclei $R_{A/D}^p(k)=R_{A/D}^n(k)\equiv R_{A/D}(k)$, whereas in  $^3$He  $R_{A/D}^p(k)\neq n_A^n(k)/n_D(k)$. ({\bf b}):
 the proton and neutron ratios in $^3$He. Wave functions as in Fig. \ref{Fig6}. (Figure reprinted from. \cite{Alvioli:2013}.
Copyright (2013) by the American Physical Society)} \label{Fig9}
\end{figure}

It  would appear from  Fig. \ref{Fig6}({\bf b}), that at $k\gtrsim 1.5-2\,fm^{-1}$ the proton
momentum distribution  in $A\geq 3$ nuclei would be nothing but  the rescaled deuteron
momentum distribution.
Such a possibility has been quantitatively investigated in Ref. \cite{Alvioli:2013}
 by plotting the  ratio $R_{A/D}^N(k)=n_A^N(k)/n_D(k)$. The results are presented  in Fig. \ref{Fig9}({\bf a}),
  which shows the proton  ratio for $A \geq 3$, and in Fig. \ref{Fig9}({\bf b}), which shows the proton and neutron ratios in $^3$He.
  The linear scale  demonstrates  that, starting from $k \gtrsim 2\,fm^{-1}$, the ratio $R_{A/D}^N(k)$ is not
   constant but appreciably increases with $k$. The reasons for such an increase are manyfold, namely \cite{Alvioli:2013}: (i) the role
 of the states $(ST)=(01)$ and $(11)$, that are missing in the deuteron; (ii) the \emph{c.m.} motion of a pair in a
 nucleus, that, unlike what happens in  the deuteron, is not zero;
 (iii) the different role played by $pp$ and $pn$ SRCs.
  In order to  better understand the last point,
 let us analyze in detail the proton and neutron momentum distributions in ${^3}$He.
\begin{figure}[tbph!]
\begin{center}
\includegraphics[scale=0.35]{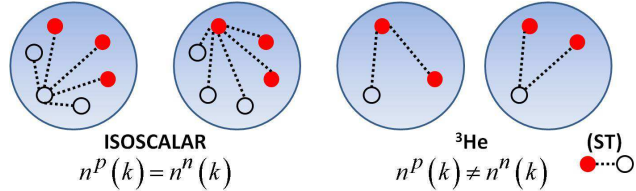}
\caption{The number of  $pn$ and $pp$ pairs  affecting  the high-momentum components of the nucleon momentum distributions.
In  isoscalar nuclei $n^p(k)=n^n(k)$,  whereas in non isoscalar nuclei, e.g. in $^3$He,
 $n^p(k)\neq n^n(k)$ because the proton and the neutron are
correlated with different nucleon pairs.(Full (open) dots denotes  protons(neutrons).}
\label{Fig10}
\end{center}
\end{figure}
\subsection{The nucleon momentum distributions in $^3$He and $^3$H}
\label{subsec:4.6}
\begin{figure}[tbph!]
\vspace{-1.5cm}
\hspace{-1.2cm}
\begin{minipage}[c]{0.95\textwidth}
\centering
\includegraphics[scale=0.45]{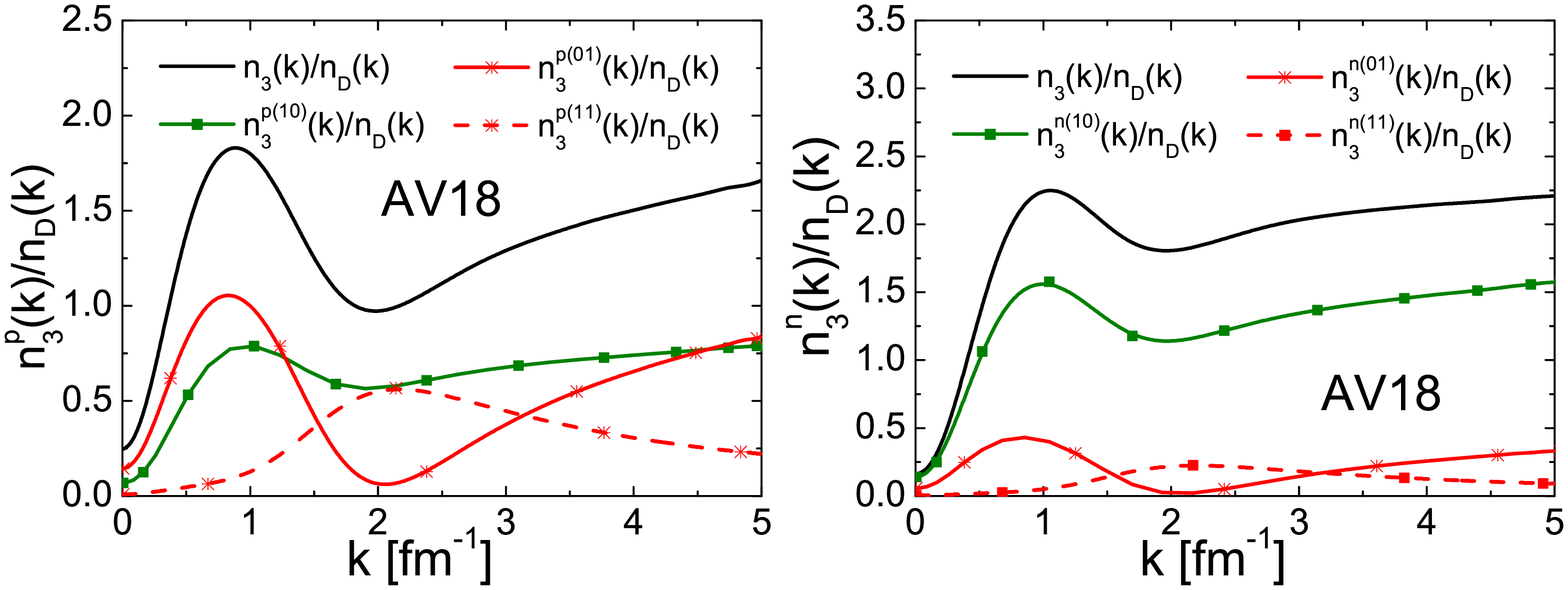}
\end{minipage}%
\vspace{-0.8cm}
\caption{The spin-isospin components of the proton ({\bf Left}) and neutron ({\bf Right}) ratios
$n_A^N(k)/n_D(k)=\sum_{ST} n_3^{N,(ST)}(k)/n_D(k)$ in $^3$He. Wave functions from Ref.\cite{Kievsky:1992um}.
(Figure reprinted from. \cite{Alvioli:2013}.
Copyright (2013) by the American Physical Society).} \label{Fig11}
\end{figure}
The different behavior of
 the proton and neutron momentum ratios shown in Fig. \ref{Fig9}({\bf a}),
can  be
 understood in terms of SRC as follows \cite{Alvioli:2013}.
  A $pn$ pair can be either in deuteron-like $(10)$ state with probability $3/4$, or in $(01)$ state, with probability
  $1/4$;  a $pp$ ($nn$) pair can only be  in $(01)$ state with probability  one. \footnote{This is strictly
  true in the independent particle picture. SRCs  change these probability according to the results
  presented in Table \ref{Table1} without, however, affecting the correctness of our argument.}
  As illustrated in the cartoon in Fig \ref{Fig10},
  in $^3$He the proton momentum distribution is affected by SRCs acting in one
 $pn$  and one $pp$ pairs; in the former pair  the deuteron-like state $(10)$ is three
 times larger than the $(01)$ state,  whereas in the latter pair the
 deuteron-like state is totally missing; on the  contrary, the neutron distribution
  is affected by SRCs acting in two  proton-neutron pairs, with a pronounced dominance of the
  deuteron-like state $(10)$; therefore,
  one expects  that
  around $k \simeq 2\,fm^{-1}$,  where
$np$ SRCs  dominate over
 $pp$ SRCs \cite{Sargsian:2005ru,Schiavilla:2006xx,Alvioli:2007zz},  $n_3^n/n_D \simeq 2$ and
  $n_3^p/n_D \simeq 1$, which   is indeed confirmed by the results presented in Fig. \ref{Fig11},
where the various spin-isospin ratios
$R_{A/D}^{N,(ST)}(k)=n_{A}^{N,(ST)}(k)/n_D(k)$ are presented.
\subsection{Experimental evidence of high-momentum components in the one-body momentum distributions}
\label{subsec:4.7}
As already pointed out in Section \ref{sec:1},  it is not the aim of the present review
to discuss  the experimental investigation
\begin{figure}[tbph!]
\hspace{-0.5cm}
\begin{minipage}[c]{0.95\textwidth}
\centering
\includegraphics[scale=0.37]{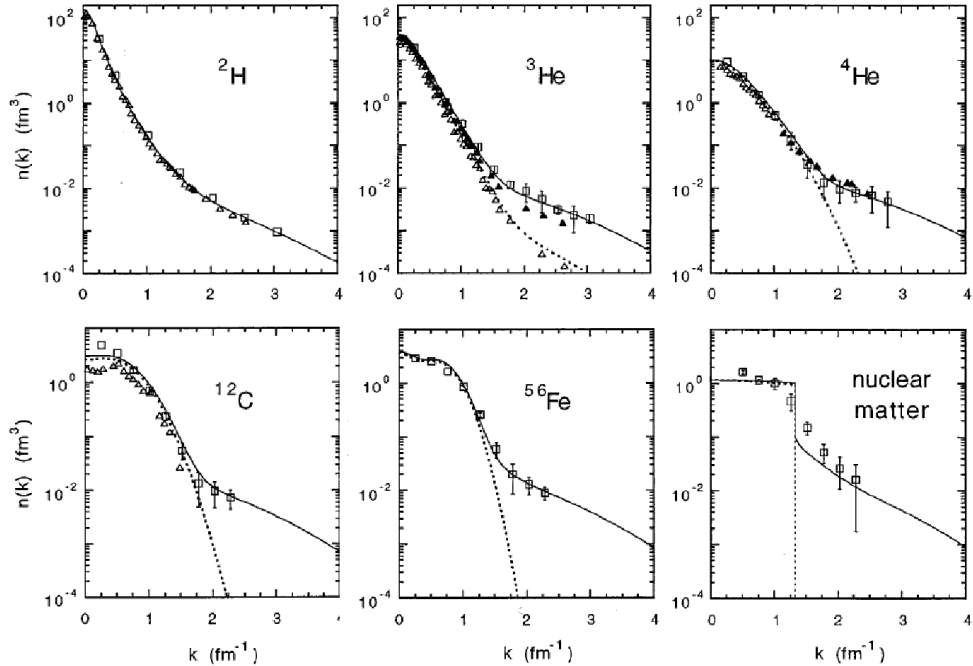}
\end{minipage}%
\vspace{-0.5cm}
\caption{The momentum distributions of several nuclei and nuclear matter  extracted from the  analysis
of inclusive,  $A(e,e')X$ (open squares),   and exclusive, $A(e,e'p)X$ (full and open triangles), cross sections. The full lines
represent the results of many-body calculations, as in the previous Figures,  and the dashed lines are MF predictions.
For references to the original experimental
and theoretical papers see Ref. \cite{CiofidegliAtti:1990rw}.(Figure reprinted from. \cite{CiofidegliAtti:1990rw}.
Copyright (1991) by Elsevier)}
\label{Fig12}
\end{figure}
and the evidence of SRCs, in particular, how the information on momentum distributions
could be  extracted from different types of measured cross sections  which might be
strongly affected by competitive  effects, like  the final state interaction (FSI) and
meson exchange currents (MEC). Nonetheless it is useful mentioning some established
evidence of high-momentum components in $n_A(k)$. To this end, we show in Fig.
\ref{Fig12} the one-body momentum distributions extracted from the exclusive, $A(e,e'p)X$, and inclusive,
 $A(e,e')X$, reactions, the latter  analyzed in terms of y-scaling \cite{CiofidegliAtti:1990rw}. The y-scaling analysis produce
 large errors, but even in the worst case, it
unambiguously demonstrates the dominant role played by SRCs in the high-momentum part of
the one-body momentum distributions. Other evidence of SRCs from inclusive electron
scattering is provided by the ratio of inclusive cross sections, e.g.
$\sigma_A(x_{Bj},Q^2)/\sigma_D(x_{Bj},Q^2)$, plotted {\it vs.} the Bjorken scaling
variable $x_{Bj}$ (see Ref. \cite{Fomin:2012,Chiara:2009} and the review
paper\cite{Arrington:2012}).
\section{Two-body momentum distributions}
\label{sec:5}
Introducing the relative and \emph{c.m.} momenta,
\bea
\Vec{k}_{rel}=\frac{1}{2}\,(\,\Vec{k}_1-\Vec{k}_2\,)\,\equiv\,\Vec{k}
\,\,\,\,\,\,\,\,\,\,\,\,\, \Vec{K}_{c.m.}=\Vec{k}_1+\Vec{k}_2 \,\equiv\,\Vec{K},
\label{Rel_C__momenta}
\eea
 the two-body momentum distribution is defined
 as follows
\bea
&&n(\Vec{k}_1,\Vec{k}_2)=n(\Vec{k},\Vec{K})=
 {n({k},{K},\Theta)} =\nonumber\\
&&=\frac{1}{(2\pi)^6}\int d\Vec{r}d\Vec{r}^\prime
d\Vec{R}d\Vec{R}^\prime\,e^{-i\,\Vec{K}\cdot(\Vec{R}-\Vec{R}^\prime)}
\,e^{-i\,\Vec{k}\cdot(\Vec{r}-\Vec{r}^\prime)}
\rho(\Vec{r},\Vec{r}^\prime;\Vec{R},\Vec{R}^\prime), \label{2BMD} \eea where
$\rho(\Vec{r},\Vec{r}^\prime;\Vec{R},\Vec{R}^\prime)$ is the non-diagonal two-body
density  (Eq. (\ref{2BND})), $k=|\Vec{k}|$, $K=|\Vec{K}|$ and $\Theta$ is  the angle
between ${\bf k}$ and ${\bf K}$. Three different types of  two-body momentum distribution
can  thus be
 considered, namely:
 \begin{enumerate}
\item the relative,  $n_{rel}(k)$,  and  \emph{c.m.}, $n_{c.m.}(K)$, momentum distributions,
i.e. Eq. (\ref{2BMD})  integrated over the \emph{c.m.} and relative momenta, respectively:
\bea
 n_{rel}(\Vec{k}) = \frac{1}{(2\pi)^3}\int n(\Vec{k},\Vec{K})\,d\Vec{K}
\,\,\,\,\,\,\,\,\,\,\,\,  n_{c.m.}(\Vec{K})=\frac{1}{(2\pi)^3}\int
n(\Vec{k},\Vec{K})\,d\Vec{k};
\label{2BMD_integrated}
\eea
\item Eq. (\ref{2BMD}) in correspondence of  $K_{c.m.}=0$, describing  back-to-back nucleons, as in the deuteron  ($\Vec{k}_2 =
-\Vec{k}_1$) :
\bea
n(\Vec{k},0)=\frac{1}{(2\pi)^6}\int d\Vec{r}d\Vec{r}^\prime\,
\,e^{-i\,\Vec{k}\cdot(\Vec{r}-\Vec{r}^\prime)}
\int d\Vec{R}d\Vec{R}^\prime \,\rho(\Vec{r},\Vec{r}^\prime;\Vec{R},\Vec{R}^\prime);
\label{2BMD_Back-to-Back}
\eea
\item the full Eq. (\ref{2BMD}) as a function of $k$,  $K$ and $\Theta$,  a quantity that  provides a three-dimensional picture of
the two-body momentum distributions.
\end{enumerate}
Hereafter, the two-body momentum distributions for a pair of nucleons $N_1N_2$ in
spin-isospin state $(ST)$ will be denoted by $n^{N_1N_2}_{(ST)}(\Vec{k},\Vec{K})$.
\begin{figure}[tbph!]
\begin{minipage}[c]{.4\textwidth}
\centering
\includegraphics[scale=0.25,angle=270]{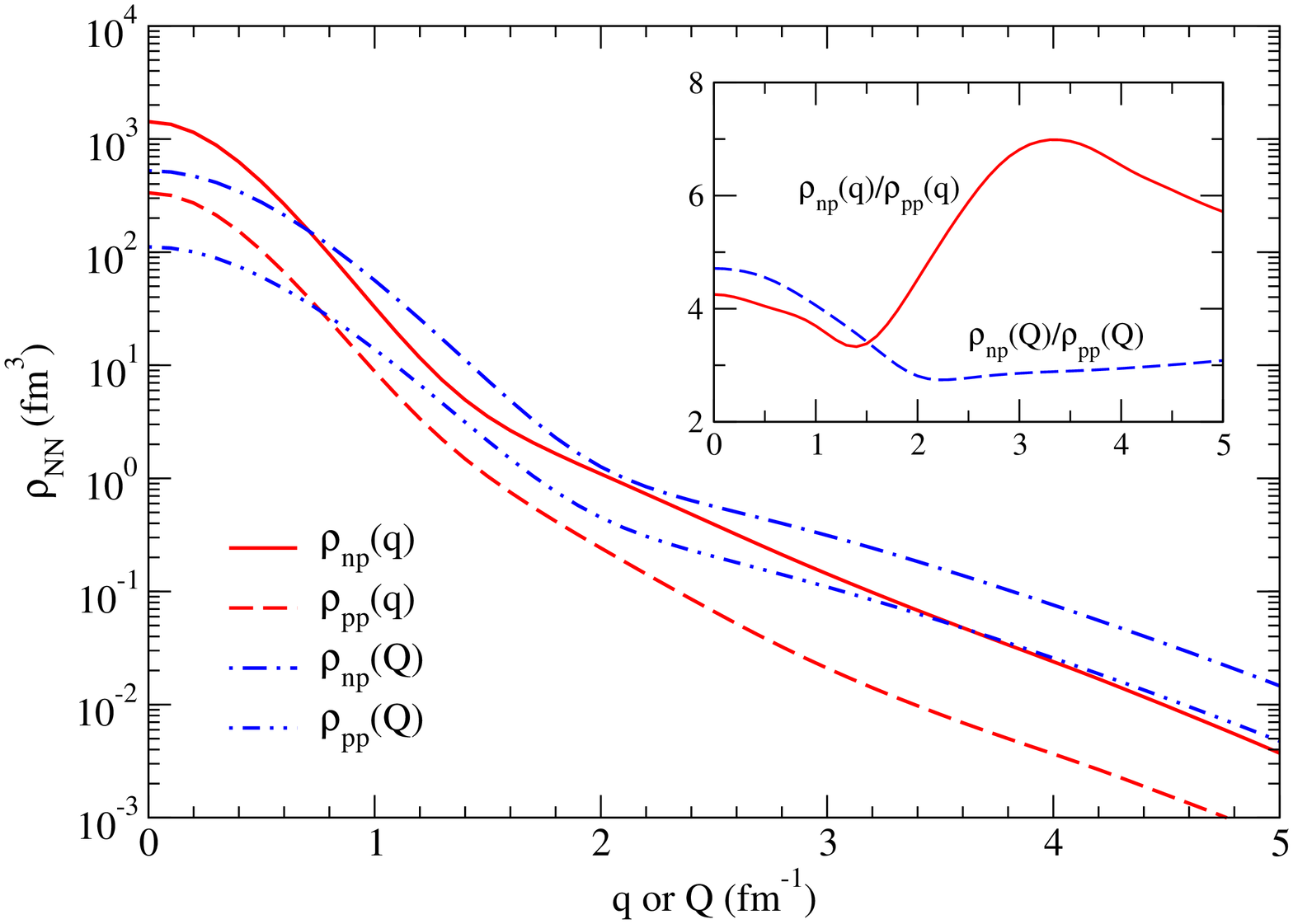}
\end{minipage}%
\hspace{1.cm}%
\begin{minipage}[c]{.4\textwidth}
\centering
\includegraphics[scale=1.1]{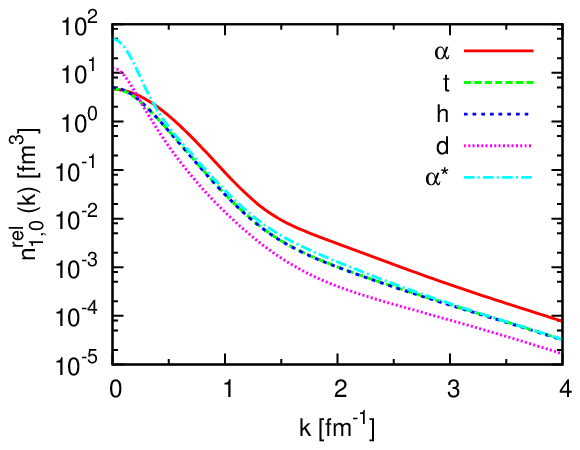}
\end{minipage}
\caption{(Color online) ({\bf Left}): the two-body momentum distribution of $np$ and $pp$
pairs in $^4$He, integrated over the the \emph{c.m.} (relative) momentum  ${K} \equiv {Q}$ ($k \equiv q$)  {\it vs.} the
relative (c.m.) momentum  (Eq. (\ref{2BMD_integrated})), with
$n_{rel}(k)\equiv\rho_{NN}(q)$ and $n_{c.m.}(K)\equiv\rho_{NN}(Q)$. The inset shows the
ratios $\rho_{np}(q)/\rho_{pp}(q)$ and $\rho_{np}(Q)/\rho_{pp}(Q)$. (Figure reprinted
from \cite{Schiavilla:2006xx}. Copyright (2007) by the American Physical Society). ({\bf
Right}): the same as in Fig. \ref{Fig13}({\bf Left}) for a pair in
 $S=1, T=0$ channel in $^2$H $\equiv$ d, $^3$He $\equiv$ h, $^3$H $\equiv$ t, $^4$He $\equiv$ $\alpha$ and
$^4$He$^*$ $\equiv \alpha^*$. (Figure reprinted from. \cite{Feldmeier:2011qy}.
Copyright (2011) by the American Physical Society).}
\label{Fig13}
\end{figure}
\subsection{The momentum distributions $n_{rel}(k)$ and  $n_{c.m.}(K)$}
\label{subsec:5.1}
Fig. \ref{Fig13}({\bf Left}) shows the relative and \emph{c.m.}  momentum distributions in $^4$He obtained in Ref.\cite{Schiavilla:2006xx}
  whereas  Fig. \ref{Fig13}({\bf Right}) shows  the
 the relative momentum distributions for  $pn$  pairs in  state $(ST)=(10)$ in
$^2$H, $^3$He, $^3$H, $^4$He and$^4$He$^*$ from Ref.\cite{Feldmeier:2011qy}.
 Both calculation are {\it ab initio} within the VMC  method with the AV18 interaction (Ref. \cite{Schiavilla:2006xx})
 and the correlated basis approach with the AV8$^{\prime}$
interaction (Ref. \cite{Feldmeier:2011qy}). The inset in Fig. \ref{Fig13} illustrates
the dominance of the tensor force acting in  $pn$  pairs:
at low momenta, the  ratio   $n_{np}(k)/n_{pp}(k)$ is mostly governed by the ratio of the $pn$ to \emph{pp} pairs, $ZN/[Z(Z-1)/2]
=2$ but starting from $k\geq 1.5\,fm^{-1}$, the ratio  sharply increases because of the action of the tensor force
in the $(10)$ channel of the $np$ pair. The results exhibited in Fig. \ref{Fig13} demonstrate the universality of
SRCs in  few-nucleon system:  at high values of $k$, the relative momentum distributions are
very similar, thanks to the universality
of the correlation hole previously discussed in Section \ref{subsec:2.3}. The universality of the integrated
momentum distributions is confirmed by the results for A=12, 16 and 40, obtained in Ref. \cite{Alvioli_1}
within the   number-conserving linked-cluster expansion  and
 the AV8$^{\prime}$ $NN$ interaction.
\subsection{The momentum distributions  $n(k_{rel},K_{c.m.}=0)$}
\label{subsec:5.2}
The momentum distribution  $n(k_{rel},K_{c.m.}=0)$ is a very important  quantity because,
when compared
 with  the deuteron momentum distribution, it
 can provide information on the short-range dynamics  of a pair of nucleons in the medium
 and possible evidence of  medium induced multi-nucleon correlations.
\begin{figure}[tbph!]
\begin{center}
\hspace{-0.5cm}
\includegraphics[scale=0.26,angle=270]{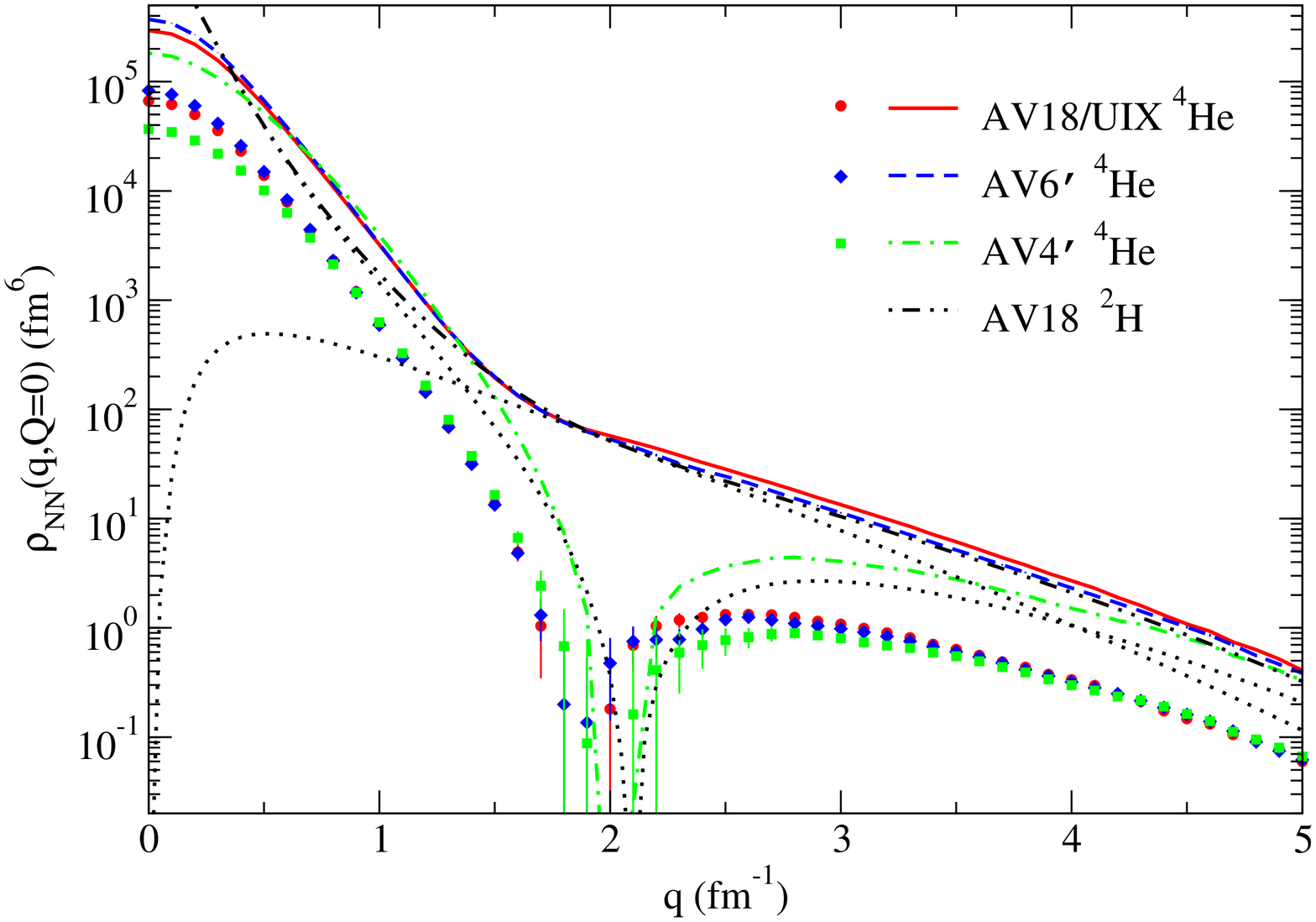}
\hspace{-0.1cm}
\includegraphics[scale=0.26,angle=270]{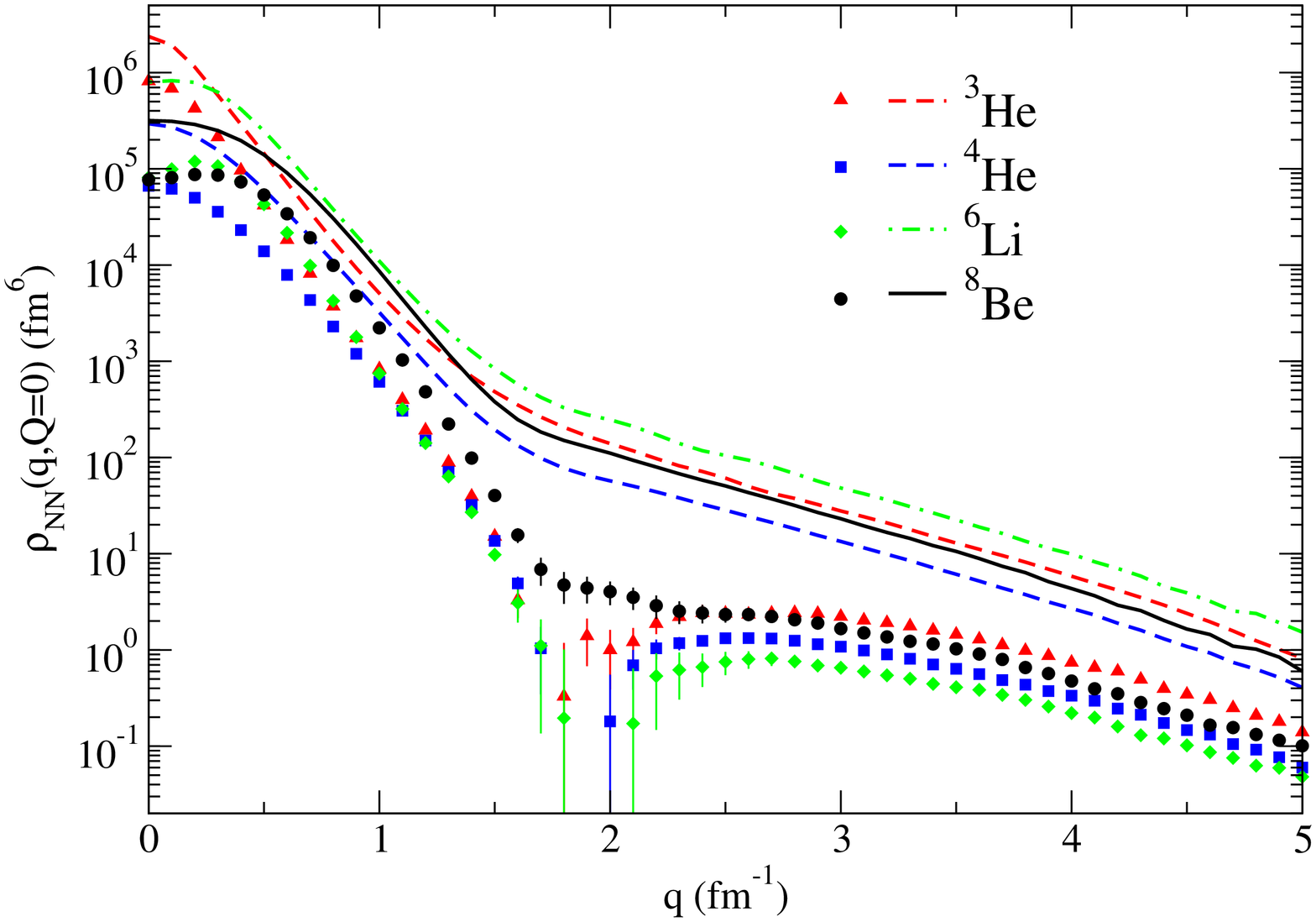}
\caption{(Color online) ({\bf Left}): the two-body momentum distributions of back-to-back
nucleons (Eq. (\ref{2BMD_Back-to-Back}) with  $K_{c.m.}\equiv Q=0$ and
 $n(k,0)\equiv \rho(q,Q=0)$) for   $np$ pairs (lines) and $pp$ pairs (symbols)
  in $^4$He, calculated with
  VMC
 wave functions and different NN interactions:
AV18 plus UIX three nucleon interaction
\cite{UIX_3NF}, AV6$^{\prime}$ \cite{AV6} and  AV4$^{\prime}$ \cite{AV6} interactions. The dotted lines denote the S and D waves
 of the deuteron corresponding to the AV18 interaction.
({\bf Right}): the same as in ({\bf Left}) but  for   $^3$He, $^4$He,
$^6$Li, and $^8$Be. AV18 interaction.(Figure reprinted from.
\cite{Schiavilla:2006xx}. Copyright (2007) by the American
Physical Society)} \label{Fig14}
\end{center}
\end{figure}
\begin{figure}[tbph!]
\hspace{-0.9cm}
\begin{minipage}[c]{0.95\textwidth}
\centering
\includegraphics[scale=0.45]{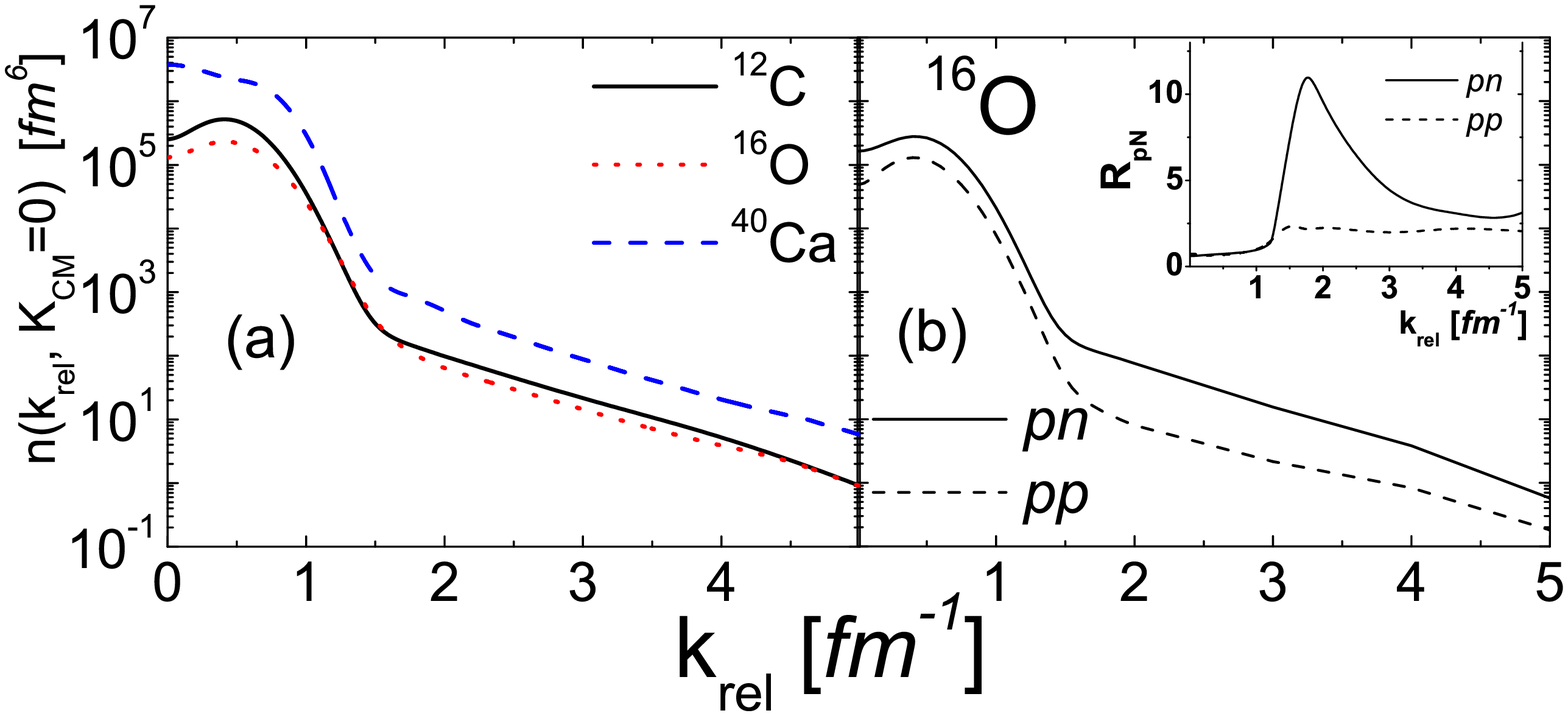}
\end{minipage}%
\vspace{-0.9cm}
\caption{(Color online) ({\bf a}): the two-nucleon momentum distribution  in
$^{12}$C, $^{16}$O and $^{40}$Ca for back-to-back nucleons (Eq. (\ref{2BMD_Back-to-Back}))
calculated within the number-conserving linked-cluster expansion of Ref. \cite{Alvioli:2005cz}
with AV8' interaction. ({\bf b}): the  back-to-back $pn$ and $pp$ momentum distributions in  $^{16}O$. The inset shows the ratio
  of the total momentum distributions to the distributions obtained by disregarding the tensor force, i.e.
  $R_{pN}= n_{pN}({k}_{rel},{K}_{c.m.}=0)/n_{pN}^{central}({k}_{rel},{K}_{c.m.}=0)$. All curves are normalization to
  the number of NN pair.
  (Figure reprinted from. \cite{Alvioli:2007zz}.
Copyright (2008) by the American Physical Society)} \label{Fig15} \vspace{-0.3cm}
\end{figure}


 The results  for   A=3, 4, 6 and 8 nuclei, obtained in Ref. \cite{Schiavilla:2006xx} within the VMC
  method using different NN interactions plus 3N forces,  are shown in Fig. \ref{Fig14},
 whereas the results  for A=12, 16 and 40,  obtained in  Ref. \cite{Alvioli:2007zz} within
  the number-conserving linked-cluster expansion  and
 the AV8$^{\prime}$ NN interaction,  are shown in Fig. \ref{Fig15}.
The results for both few-nucleon systems and complex nuclei clearly show that:
 (i)
 the 3NF, which is essential to produce the correct binding energy of few-nucleon systems,
 appears to have tiny effects  on the high-momentum components (Fig. \ref{Fig14}({\bf Left})), which is not surprising, in view of
 its long-range character;
 (ii) the universality of the relative momentum
 distributions, resulting from the universality of SRCs, is evident from Fig. \ref{Fig14}({\bf Right})
 and Fig. \ref{Fig15}({\bf a}): in the range $3\leq A \leq 40$  and $k \geq \,2\,fm^{-1}$
 a clear A-independence of the high relative-momentum behavior is exhibited; (iii) the results presented in Figs. \ref{Fig13}({\bf a}) and \ref{Fig15}({\bf b})
 demonstrate  the tensor dominance both in few-nucleon systems and  complex nuclei; (iv)  Fig.
 \ref{Fig14}({\bf Left}) and \ref{Fig15}({\bf b})  shows that  at high values of $k_{rel}$  the momentum distributions of deuteron
 and complex nucleus are very similar. This similarity is better illustrated
 in the next Section where the ratio $R_{A/D}(k, K_{c.m.}=0)=n_A^{pn}(k_{rel}, K_{c.m.}=0)/n_D(k)$ is presented for few-nucleon systems and
 complex nuclei.

\subsection{The momentum distributions  $n(k_{rel},K_{c.m.},\Theta)$}
\label{subsec:5.3}
\begin{figure}[tbph!]
\hspace{-1.2cm}
\begin{minipage}[c]{0.95\textwidth}
\centering
\vspace{-1.0cm}
\includegraphics[scale=0.48]{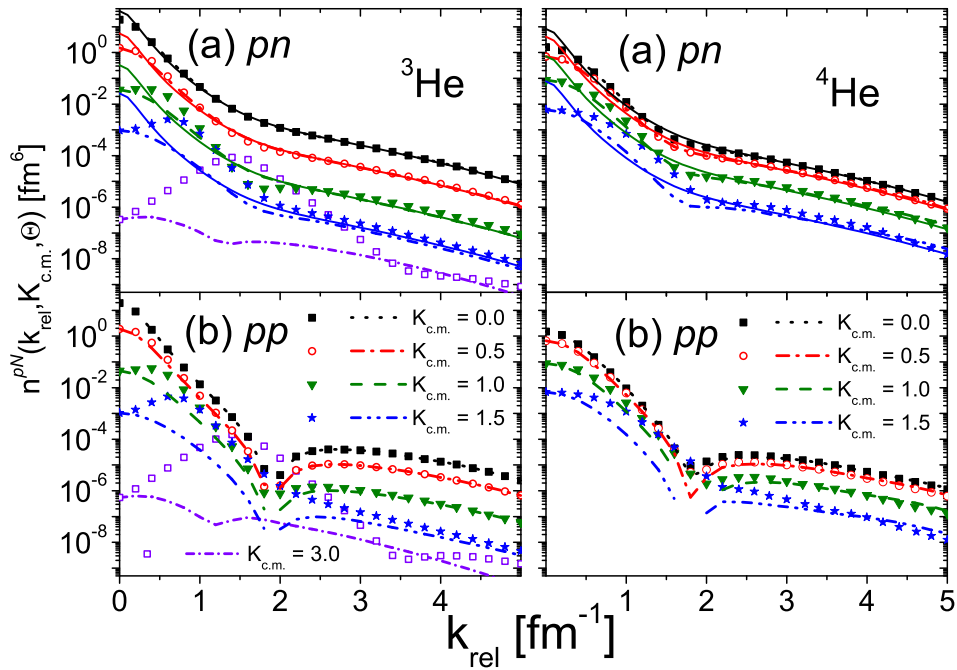}
\end{minipage}%
\vspace{-0.5cm}
\caption{(Color online) ({\bf Left}): the  two-body momentum
  distributions   of   $pn$ ({\bf a})   and $pp$ ({\bf b}) pairs in $^3$He normalized to unity, {\it vs.} the  relative momentum
  $k_{rel}$,  for fixed values of the
   c.m. momentum $K_{c.m.}$ and two orientations of the momenta, namely
   ${\bf k}_{rel}
    ||{\bf K}_{c.m.}$ ({\it broken curves}) and ${\bf k}_{rel} \perp {\bf
    K}_{c.m.}$ ({\it symbols}). The continuous curves for the $pn$ pair represents the deuteron momentum distribution
    rescaled by the c.m. momentum distribution
    $n_{c.m.}^{pn}(K_{c.m.})= \int n^{pn}(\Vec{k}_{rel},\Vec{K}_{c.m.})\, d\Vec{k}_{rel}$ (see Eq. (\ref{2BMD})).
  $^3$He  wave function from Ref. \cite{Kievsky:1992um} and
  $AV18$ interaction \cite{AV18}. ({\bf Right}): the same as in ({\bf Left}) but for $^4He$. Correlated variational  wave function  from
  \cite{Akaishi:1988} and
    $AV8'$ interaction \cite{Pudliner:1997ck}.(Figure reprinted from. \cite{Alvioli:2012aa}.
Copyright (2012) by the American Physical Society)}
    \label{Fig16}
    \end{figure}

\begin{figure}[tbph!]
\centering
\includegraphics[scale=0.30]{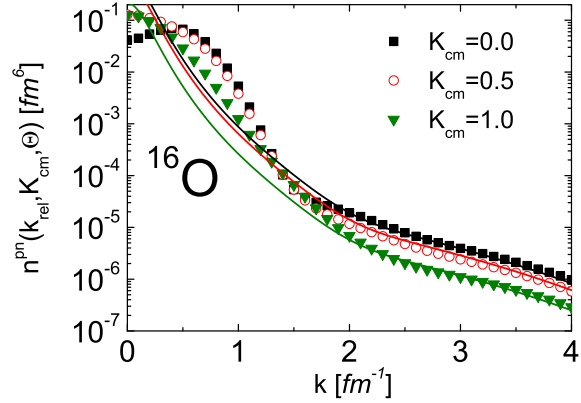}
  \vskip -0.5cm
  \caption{ The two-body proton-neutron momentum distributions
 (Eq. (\ref{2BMD})) in
  $^{16}$O  for three values of $K_{c.m.}$ and $\Theta=0$ (symbols). The continuous lines represent the deuteron
  momentum distributions rescaled by the c.m. momentum distribution of the pair calculated at the proper value of $K_{cm}$
  (see Fig. \ref{Fig18}({\bf b})). Wave function from the
  number-conserved linked-cluster expansion of Ref. \cite{Alvioli:2005cz}. AV8$^{\prime}$ NN interaction.
  (Figure adapted from  \cite{Alvioli:2007zz}.
Copyright (2008) by the American Physical Society)}
 \label{Fig17}
\end{figure}

The knowledge of  $n(k_{rel},K_{c.m.},\Theta)$  provides information on
 the three-dimensional picture of
 the two-nucleon momentum distribution. In this connection, it has to be stressed that
  the  independence of $n(k_{rel},K_{c.m.},\Theta)$
 upon the angle $\Theta$  is evidence of the factorization of
the distributions in  variables $k_{\text{rel}}$ and $K_{\text{c.m.}}$ \cite{Alvioli:2012aa,CKMS_fewbody}, i.e.
$n^{NN}({k}_{\text{rel}}, {K}_{\text{c.m.}},\Theta) \Rightarrow \phi(k_{\text{rel}})\chi({K}_{\text{c.m.}})$
where, for the time being, $\phi(k_{\text{rel}})$ and $\chi({K}_{\text{c.m.}})$ denote
two generic functions of $k_{rel}$ and $K_{c.m.}$ . The $pn$ and $pp$ two-body momentum
distributions  $n(k_{rel},K_{c.m.},\Theta)$ in few-nucleon systems \cite{Alvioli:2012aa}
and complex nuclei \cite{Alvioli_1} have been calculated with realistic wave functions.
The results for $^{3}$He and  $^{4}$He obtained with {\it ab initio} wave functions
\cite{Kievsky:1992um,Akaishi:1988} corresponding to the $AV18$ \cite{AV18} and $AV8'$
\cite{Pudliner:1997ck} interactions  are shown in Fig. \ref{Fig16}  {\it vs.}
$k_{\text{rel}}$, in correspondence of several values of $K_{\text{c.m.}}$ and two values
of $\Theta$. The results for $^{16}$O are shown in Fig. \ref{Fig17}.
\begin{figure}[tbph!]
\hspace{-0.5cm}%
\begin{minipage}[c]{.4\textwidth}
\centering
\includegraphics[scale=0.25]{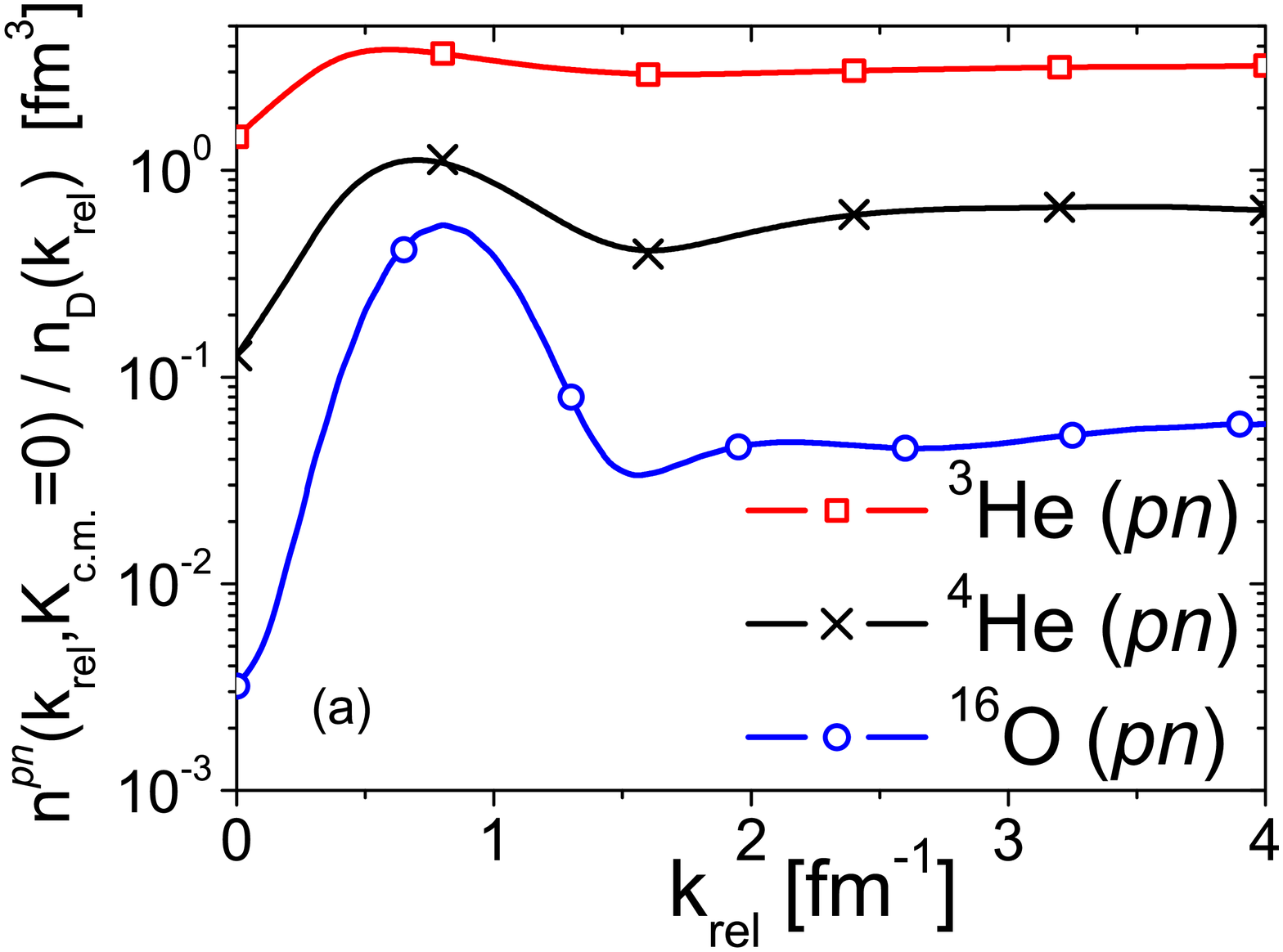}
\end{minipage}%
\hspace{1.5cm}%
\begin{minipage}[c]{.4\textwidth}
\centering
\includegraphics[scale=0.25]{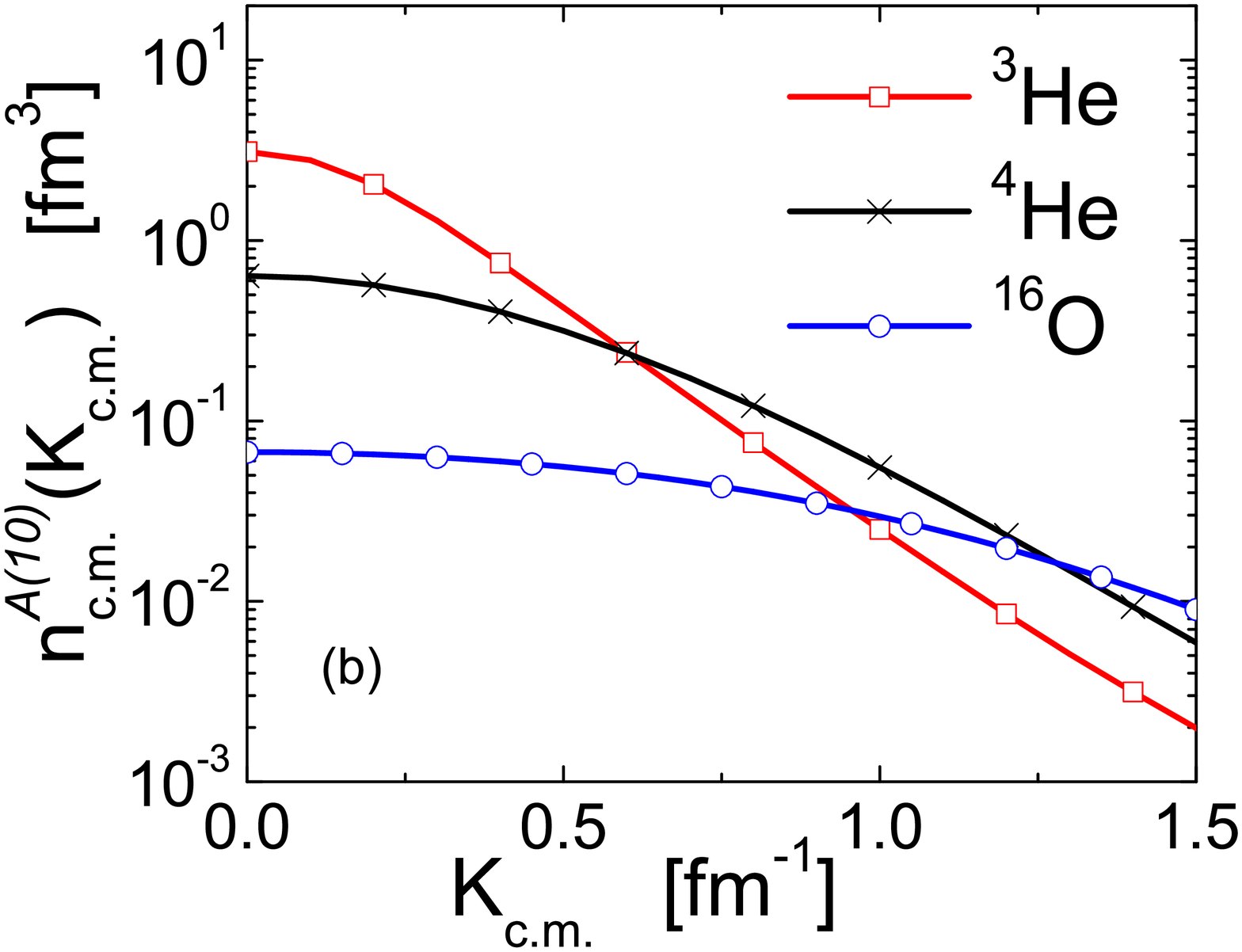}
\end{minipage}
\vspace{-0.5cm}
 \caption{{(\bf a)}: the ratio of the $pn$  momentum
  distributions  for back-to-back nucleons $n^{pn}(k_{\text{rel}},K_{\text{c.m.}}=0)$
  in $^3$He,$^4$He, and  $^{16}$O shown in
  Figs. \ref{Fig16} and \ref{Fig17},
 to the deuteron momentum distribution $n_D(k_{\text{rel}})$ (full
 lines).
    The different magnitudes of the ratio for  the  three  nuclei is due to the
    different values of the
     c.m. momentum distribution at $K_{\text{c.m.}}=0$ shown in Fig.\ref{Fig18}({\bf b}). ({\textbf b}): the \emph{c.m.} momentum distribution
     in $^3$He, $^4He$, and $^{16}$O. (Figure reprinted from. \cite{Alvioli:2012aa,Alvioli:2013}
Copyright (2012,2013) by the American Physical Society)}
\label{Fig18}
\end{figure}
Apart from a different overall normalization,  the results for few-nucleon systems  at $K_{\text{c.m.}}=0$
fully agree with the ones
  of Ref. \cite{Schiavilla:2006xx}. The peculiar and systematic results
  of these calculations can be summarized as follows: (i) with  increasing values of the c.m.  momentum,
 the high relative momentum part of the
distributions strongly decreases; (ii)  starting from a given value of
$k_{\text{rel}}$, being $k_{\text{rel}}\simeq 1.5$ fm$^{-1}$ when $K_{\text{c.m.}}=0$  and assuming
 increasing values with increasing values of
 $K_{\text{c.m.}}$, the $pn$ distribution
 changes its slope and becomes close to the
deuteron distribution. In particular,   in the  region (${k}_{\text{rel}}\gtrsim 2
\,\text{fm}^{-1},{K}_{\text{c.m.}}\lesssim 1$ fm$^{-1}$), $n^{pn}$ becomes
$\Theta$-independent\footnote{Such an independence has been
checked in a wide range of angles}, assuming the form
$n^{pn}({k}_{\text{rel}},{K}_{\text{c.m.}},\Theta)\simeq n_D(k_{\text{rel}})n_{\text{c.m.}}^{pn}({K}_{\text{c.m.}})$,
 where $n_D(k_{\text{rel}})$ is
the deuteron momentum distribution, and $n_{\text{c.m.}}^{pn}({K}_{\text{c.m.}})$ describes the c.m. motion of the pair
and provides the
 A-dependence of $n^{pn}({k}_{\text{rel}},{K}_{\text{c.m.}})$. The factorized property, that charaterizes  also  complex nuclei,  as shown in the case of $^{16}$O in Fig.
\ref{Fig17}, represents a rigorous many-body demonstration that when the relative
momentum of the \emph{pn} pair is high, and, at the same time, the c.m. momentum is low,
the two-body momentum distribution factorizes; (iii)  when the \emph{c.m.} momentum is of
the same order of the (high)
 relative momentum, more than two particles can be locally
correlated, with a resulting strong dependence upon the angle and the breaking down of
factorization, as clearly appears  in Fig. \ref{Fig16} for $K_{\text{c.m.}}= 3$
fm$^{-1}$. These feature are common to both few-nucleon systems and complex nuclei. A
better evidence on
 the factorized behavior  of the two-body momentum distributions for
  $pn$ pairs
 can be obtained by considering the ratio $R^{pn}(k)=
n^{pn}(k_{\text{rel}},0)/n_D(k_{\text{rel}})$,  which is presented in  Fig. \ref{Fig18}({\bf a}).
The constant value exhibited by the ratio at $k_{\text{rel}}\gtrsim 1.5$
fm$^{-1}$ is unquestionable evidence that in this region the dependence upon
$k_{\text{rel}}$ of the two-body momentum distribution $n^{pn}(k_{\text{rel}},0)$ is the
same as in the deuteron. As for the different magnitudes of the ratio for different nuclei,
this is  governed by the c.m. motion distribution of the pair,
which is illustrated in Fig. \ref{Fig18}({\bf b}).
 It can be seen that  the difference in magnitude of the ratios   in the
region $k_{\text{rel}} \gtrsim 1.5$ fm$^{-1}$ is governed by exactly the difference between the
values of the \emph{c.m.} momentum distributions at $K_{\text{c.m.}}=0$. The more rapid fall off
of the  \emph{c.m.} momentum distributions of $^3$He, is due to the weak binding of this
nucleus, leading, with respect to the $^4$He  and $^{16}$O, to the wider separation of
the curves corresponding to various values of $K_{\text{c.m.}}$ presented in Figs.
\ref{Fig16} and \ref{Fig18}({\bf a}).  For nuclei with $A \geq 4$ and $K_{\text{c.m.}}
\lesssim 1.0-1.5$ fm$^{-1}$,  the \emph{c.m.} distribution can be associated to the average
kinetic energy $<T>_{SM}$ of a pair moving in the
 mean field with a  Gaussian distribution , $n_{c.m.}(K)\propto \exp\{-\alpha K_{c.m.}^2\}$ with
 $\alpha=3/[2<K_{c.m.}^2>]=[3(A-1)]/[4(A-2)m_N<T_{SM}>]$ as suggested in  Ref.
\cite{CiofidegliAtti:1995qe} and in agreement with
the experimental finding \cite{Tang:2002ww} for
$^{12}$C.
\section{Nucleon momentum distributions, spectral functions and SRCs}
\label{sec:6}
\begin{figure}[tbph!]
\hspace{-0.3cm}%
\begin{minipage}[c]{.4\textwidth}
\centering
\includegraphics[scale=0.55]{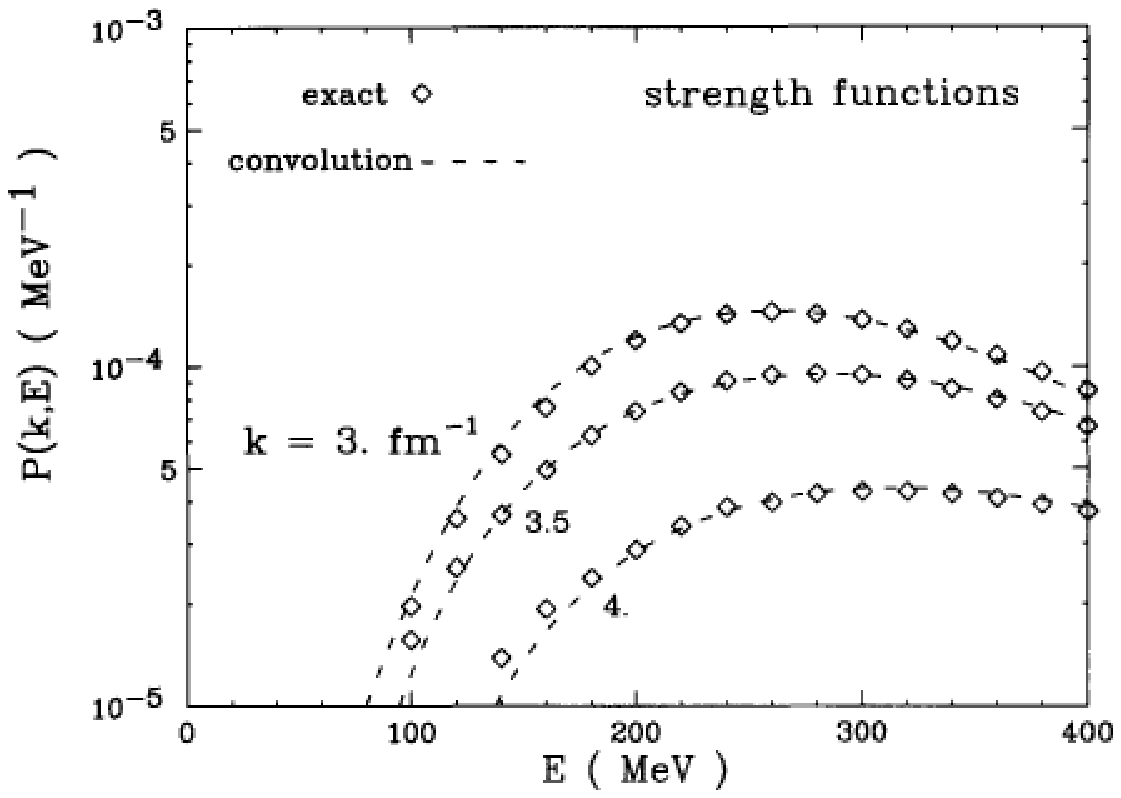}
\end{minipage}%
\hspace{1.5cm}%
\begin{minipage}[c]{.4\textwidth}
\centering
\includegraphics[scale=0.26]{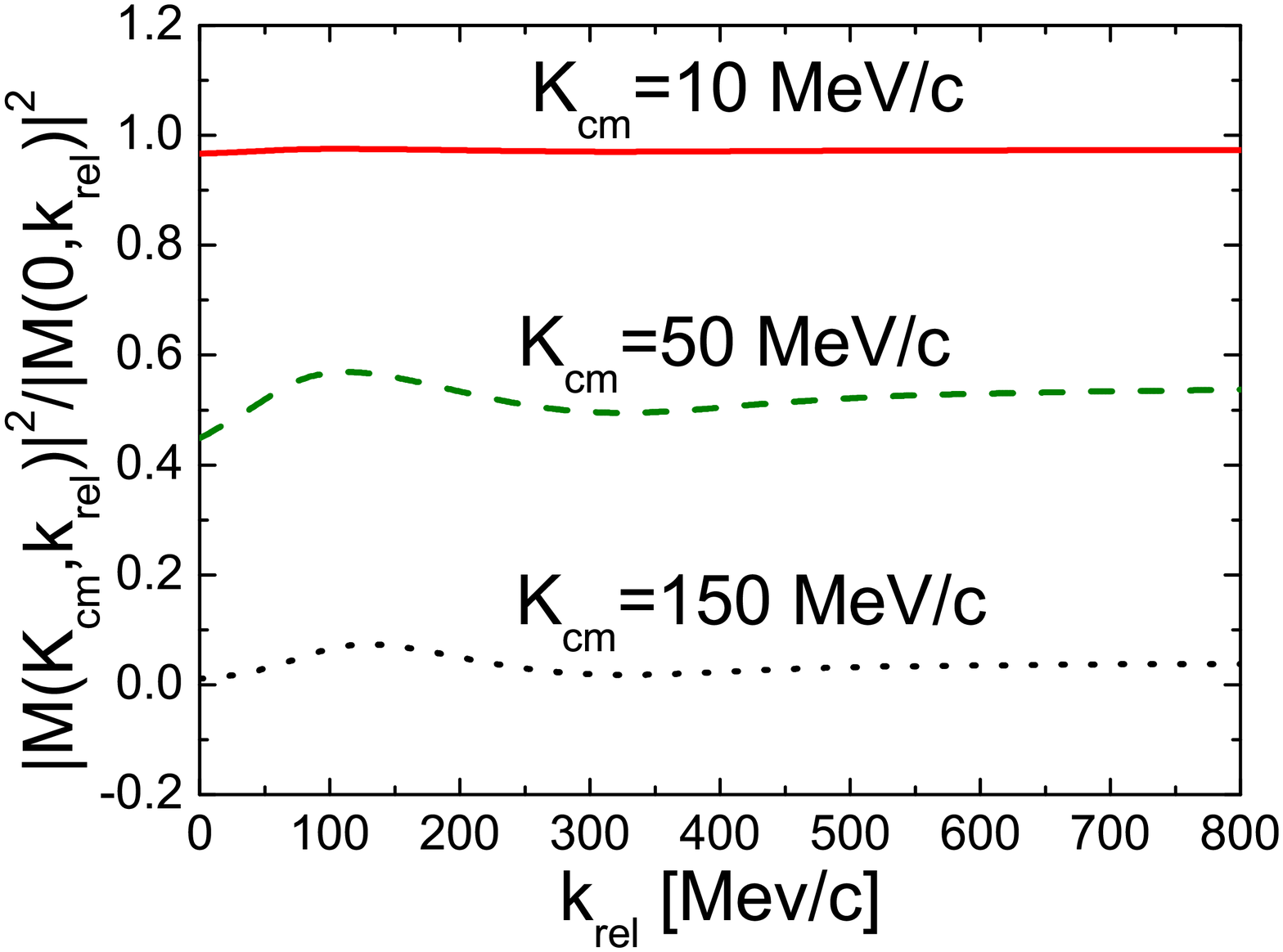}
\end{minipage}
\vspace{-0.5cm}
 \caption{({\bf Left}): the exact BBG nuclear matter spectral function ({\it exact}) {\it vs} E in correspondence of
 three values of k compared with the BBG
convolution model Eq.(\ref{Convolution_NM})({\it convolution}). (Figure reprinted from
\cite{Baldo:1996zz}. Copyright (1996) by Elsevier). ({\bf Right}): the ratio R (Eq. (\ref{Ratio_Phi})) in $^3$He. Three-nucleon wave functions from Ref.
\cite{Kievsky:1992um}. AV18 interaction \cite{AV18}.(Figure reprinted from. \cite{CKMS_fewbody}.
Copyright (2011) by Springer \& Verlag).}
\label{Fig19}
\end{figure}

 Although the momentum distribution is not an observable,
 it is undisputable that it  can play
 a role in particular scattering processes, that, at the same time, can also be
 influenced  by other phenomena which could  mask the effects generated by  the momentum distributions.
 To clarify this point, let us consider the process $A(e,e'N)X$ in the Plane Wave Impulse Approximation (PWIA), i.e.
  when, in the initial state, an electron is impinging on
 nucleus A and, in the final state, the
 scattered electron and a nucleon N are detected in coincidence  and  the nucleus $X=(A-1)$ is left in the
 energy state $E_{A-1}^f$; in the
 simplified assumption that  the detected nucleon was knocked out by a direct interaction $\gamma^*N$  and  left the
 nucleus with
 momentum ${\bf p}_N$  without interacting with the medium,
 the measurable missing momentum ${\bf p}_m = {\bf q} -{\bf p}_N$  and  energy $E_m= \nu-T_N-T_{A-1}$
  represent, respectively,
 the momentum of the nucleon before interaction ${\bf k}_1=-\bf p_m$ and the  intrinsic excitation energy of
$E_{A-1}^*$ of $(A-1)$. As is well known, even within such a severe  approximation
 the cross section of the process
 is not proportional to the momentum distribution but to another quantity, the Spectral Function $S_A({\bf k}_1,E)$
 representing the
 joint probability that when a nucleon with momentum ${\bf k}_1=-{\bf p}_m$ is removed
 instantaneously
 from the ground state of the nucleus A, the nucleus $(A-1)$ is left in
 the excited state
 $E_{A-1}^*=E-E_{min}$, where $E$ is the removal energy and $E_{min}= M_{A-1}+m_N-M_A$.
 The spectral function has  the following form (from now-on spin indexes will be omitted for ease
 of presentation)
\bea
 &&\hspace{-0.3cm}\label{1BSF2}S_A({\Vec k}_1,E)=<\Psi_{0}^A|a_{{\bf k}_1}^{\dag}
 \delta(E-\hat H+E_A)a_{{\bf k}_1} |\Psi_{0}^A>=\nonumber\\
&&\hspace{-0.3cm}=\sum_{f} \Big|\int e^{i{\bk}_1\cdot{\br}_1} d\,{\br}_1 \int
\,\Psi_{f}^{(A-1)\,*}(\{{\Vec r}\}_{A-1}) \Psi_{0}^A({\Vec r}_1,\{{\Vec r}\}_{A-1})\prod_{i=2}^A d\,{\br}_i \Big|^2\delta(E-E_{A-1}^f +E_A)\nonumber\\
\label{1BSF_2}
 &&\hspace{-0.3cm}=S_{gr(0)}({\bk}_1,E_{gr(0)}) + S_{ex(1)}({\bk}_1,E_{ex(1)})
\eea
 where  $E_{A-1}^f=E_{A-1}+E_{A-1}^*$, $E_A$ and $E_{A-1}$ denote the ground-state energies of initial and final nuclei,
 and $a_{{\bk}_1}^{\dag}
 (a_{{\bk}_1})$ is a creation (annihilation) operator. The two contributions, as in the
 case of the momentum distributions (cf. Eqs. (\ref{nsplit}) and (\ref{nsplit1})), arise from different final states of the system $(A-1)$, with
 $S_{ex(1)}({\bk}_1,E)$ governed by  SRCs.
Summing over the complete set of final states in Eq. (\ref{1BSF_2}) it is easy to
obtain the {\it momentum sum rule}
\bea
 n_A({\Vec k}_1)= \int_{0}^{\infty} S_A({\Vec k}_1,E)\,d\,E .
\label{Int_spec}
 \eea
 Eq. (\ref{Int_spec}) clearly shows that  the extraction of the momentum distribution from the experimental data
 implies a difficult  integration over the full range
 of discrete and continuum excitation spectra of the residual nucleus $(A-1)$,
 up to very high values of $E_{A-1}^*$, particularly in the interesting region of high values of $k$
 (see Ref. \cite{CPS_PkE_momdis,Ciofi_Liuti,Ciofi_Liuti_Simula}).
 Moreover, exact many-body spectral functions
exist only for the three-nucleon system \cite{Dieperink,CPS_Spectral,Sauer_Spectral},
(the complete set of final state is known), and for  nuclear matter
\cite{Omar_Spectral,Dickoff}, whereas for complex nuclei only model spectral functions
have been developed, either  within the local density approximation \cite{Omar_LDA}, or
within the convolution model  of Ref. \cite{CiofidegliAtti:1995qe}. The latter, which is
aimed at describing the spectral function in the region of 2N SRCs, naturally arises from
the behavior of the high-momentum part of the  two-body momentum distributions described
in the previous Sections. As a matter of fact, we have seen that at large values of
$k_{rel}$ and small values of $K_{c.m.}$ the following relation holds
\bea
n^{pn}({\bf k}_{rel}, {\bf K}_{c.m.}) \simeq
n^{pn}({k}_{rel},{K}_{c.m.}) \simeq n_{D}({k}_{rel})
n_{c.m.}({K}_{c.m.}).
\label{Factorization}
\eea
From momentum conservation,   ${\bf k}_1 + {\bf k}_2- {\bf K}_{c.m.}=0$, {${\bf
k}_{rel}={\bf k}_1-{\bf K}_{c.m.}/2$}, one has
\bea n_A(k_1)\simeq\int \,n_{D}(|\Vec{k}_1-\frac{\Vec{K}_{c.m.}}{2}|)
n_{c.m.}({K}_{c.m.}) \,d \,\Vec{K}_{c.m.} =\int\, S_A(k_1,E)\, d\,E,
\label{New_enne}
 \eea
\noindent and assuming that the  high values of the excitation  energy $E_{A-1}^*$
are essentially  given by the relative  motion of nucleon "2" and nucleus
$(A-2)$, one obtains
\bea
 S_A(k_1,E)&\simeq&
\int \,n_{D}(| \Vec{k}_1-\frac{\Vec{K}_{c.m.}^{N}}
{2}|)
n_{c.m.}^{N}({K}_{c.m.}) d \,\Vec{K}_{c.m.}\times\nonumber\\
&\times& \delta \left( E-E_{th}^{(2)} -\frac{A-2}{2m_N(A-1)}\left[ \Vec{k}_1
-\frac{A-1}{A-2}\Vec{K}_{c.m.} \right]^{2}\right),
 \label{Spectrl_Fun}
\eea
where $E_{th}^{(2)}$ is the two-body threshold. Eq. (\ref{Spectrl_Fun}) has been first obtained in Ref. \cite{CiofidegliAtti:1995qe},
within several phenomenological assumptions,  whose physical correctness are now justified by
the many-body calculation of the momentum distributions. A convolution formula for the correlated
part of the spectral function
 has also been shown  to result from
Brueckner-Bethe-Goldstone theory  of nuclear matter, where the spectral function
corresponding to the nucleon self-energy $M(k,E)=V(k,S)+iW(k,E)$ is obtained from the
single particle Green function ${\mathcal G}$ in the following form \cite{Dickhoff:book}
\bea S_A(k,E)=-\frac{1}{\pi}Im {\mathcal G}(k,E)=\frac{1}{\pi}
\frac{W(k,E)}{{(-E-k^2/2m_N-V(k,E))}^2 + W(k,E)^2} \label{BBG} \eea
which, at $E+\frac{k^2}{2m_N}>> |V(k,E)|;|W(k,E)|$, can be approximated by
the following
convolution integral\cite{Baldo:1996zz}
\bea
S_A(k_1,E)&=&\frac{\pi^2\,\rho}{16} \int\frac{d^3\,{\bf K}_{c.m.}}{(2\pi)^3} n_{rel}
(|{\bf k}_{1}-\frac{1}{2}{\bf K}_{c.m.}|)n_{c.m.}^{FG}(K_{c.m.})\times\nonumber\\
&\times&\delta\left ( E-E_{thr}^{(2)}-\frac{1}{2\,m_N}{({\bf K}_{c.m.}-{\bf k}_1)}^2
\right). \label{Convolution_NM} \eea
Here $\rho$ is the density of nuclear matter, $n_{c.m.}^{FG}$  the Fermi gas distribution and $n_{rel}$ the
spin-isospin averaged two-body relative
 momentum distribution in nuclear matter. Eq. (\ref{Convolution_NM}) in the region $E \simeq E_{thr}^{(2)}+ k^2/(2m_N)$
 agrees very well with the exact BBG spectral function, as shown in
  Fig. \ref{Fig19}({\bf a}).
  Further confirmation
of the convolution model, resulting from the factorization property of $n(\Vec{k}_{rel},\Vec{K}_{c.m.})$,
 has been recently
provided \cite{CKMS_fewbody} by the analysis of the behavior of  {\it ab initio}  three-nucleon ground-state wave functions
$\Psi_0$ in momentum space,
by considering  the following ratio
\bea R=\frac{|\Psi_0({\bf K}_{cm}, {\bf k}_{rel})|^2}{|\Psi_0({\bf K}_{cm}=0, {\bf
k}_{rel})|^2}, \label{Ratio_Phi} \eea
{\it vs.} $|\Vec{k_{rel}}|$ at constant values of $|\Vec{K_{c.m.}}|$ . If factorization  of $\Psi_0$ holds, i.e. $|\Psi_0({\bf K}_{cm}, {\bf k}_{rel})|^2 \simeq n_{c.m.}(K_{c.m.})n_{rel}(k_{rel})$
the ratio becomes
 \bea R\simeq\frac{n_{c.m.}(K_{c.m.})}{n_{c.m.}(K_{c.m.}=0)}= constant.
\label{Ratio_Phi_fact}
\eea
It can be seen from
Fig. \ref{Fig19}({\bf b})  that factorization indeed occurs starting, as expected,  at
a value of  $k_{rel}$ which increases with increasing values of $K_{c.m.}$, in agreement with the results presented in
Fig. \ref{Fig16}. The magnitudes of the curves in Fig. \ref{Fig19}({\bf b})
agree with the behavior of $n_{c.m.}(K_{c.m.})$ presented in Fig. \ref{Fig18}.

The most interesting quantity, as far as SRCs are concerned, is the two-nucleon momentum distributions, that in PWIA might
in principle be extracted
from the $A(e,e'2N)X$ process, when two nucleons are knocked out from the nucleus A and are detected with momenta
${\bf p}_1$ and ${\bf p}_2$ in coincidence with the scattered electron, with the nucleus $(A-2)$  left
in the energy state $E_{A-2}^f$. The measurable missing momentum and energy
are in this case the ${\bf p}_m = {\bf q} - {\bf p}_1 -{\bf p}_2$ and $E_m = \nu -T_{p_1} -T_{p_2} -T_{A-2}=E_{A-2}^*$.
Assuming that the virtual photon has interacted  with one nucleon (the fast one) of a correlated pair, with the second
nucleon (the recoiling one) being
 emitted because of momentum conservation, the cross section will depend upon the two-nucleon spectral function
\bea
 &&\hspace{-0.9cm}S_A^{N_1N_2}({\Vec k}_1,{\Vec k}_2,E)=<\Psi_{0}|a_{{\bf k}_1}^{\dag}a_{{\bf k}_2}^{\dag}
 \delta(E-\hat H+E_A)
 a_{{\bf k}_2} a_{{\bf k}_1}|\Psi_{0}>=\nonumber\\
 &&\hspace{-0.9cm}\sum_{f} \Big|\int e^{-i{\bk}_1\cdot{\br}_1-i{\bk}_2\cdot{\br}_2} d\,{\br}_1 d\,{\br}_2
  \langle\Psi_{f}^{(A-2)*}(\{{\bf r}\}_{A-2})|\Psi_{0}({\bf r}_1,{\bf r}_2\{{\bf r}\}_{A-2}) \rangle \times\nonumber\\
  &&\hspace{-0.9cm}\times\delta(E-E_{A-2}^f +E_A).
 \label{2BSF_2}
\eea
Summing over the final states of $(A-2)$  the   {\it
two-nucleon momentum sum rule}
 \bea n_A^{N_1N_2}({\Vec k}_1,{\Vec k}_2)=\int
d\,E\,S_A^{N_1N_2}({\Vec k}_1,{\Vec k}_2,E)=n({ k}_{rel},{ K}_{c.m.}, \Theta)
\label{2N_SUMRULE}
\eea
is obtained. The two-nucleon Spectral Functions has been obtained
within many body theories in Refs. \cite{Dickoff:1989} for finite nuclei,  in Ref. \cite{Omar:2Nspectral} for nuclear matter, and
in Ref. \cite{Ciofi_Kaptari} for $^3$He \footnote{ Eq. (\ref{2BSF_2}) has been called {\it vector spectral function}
 in Ref. \cite{Ciofi_Kaptari}
whereas
a similar quantity has been called {\it decay function} in Ref. \cite{Arrington:2012}}.
In the past, the process $A(e,e'N_1N_2)X$ has been intensively investigated theoretically (see
e.g. \cite{Ryckebusch:2N_emission,Giusti:2N_emission}, and references therein quoted)
and experimentally (see e.g. \cite{Groep:2001} and references therein quoted)  but the experimental  data were plagued by
MEC and FSI and other competing effects and  no conclusive quantitative
information on SRCs could be obtained  (for a critical discussion of this topic see \cite
{Arrington:2012}). Recently, however, high moment transfer  experiments have been performed on
$^{12}$C and $^4$He  \cite{Tang:2002ww,Piasetzky:2006ai,Shneor:2007,Subedi:2008} that
 allowed one to detect a "fast" proton  with momentum ${\bf p}_1$, identified as the member of a correlated pair  kicked out
 by the high energy projectile,   and a  "slow" (or "recoil")
 nucleon (a proton or a neutron) with momentum ${\bf p}_2$, assumed to be the one emitted by momentum conservation in the correlated pair.
 By assuming the validity of the  PWIA, which implies that
 ${\bf p}_1= {\bf k}_1+{\bf q}$,  ${\bf p}_2= {\bf k}_2$  and
 ${\bf P}_{mis}= -({\bf k}_1+{\bf k}_2)={\bf K}_{c.m.}$,  it is possible
 to reconstruct the momentum ${\bf k}_1$ that
 the struck nucleon had before interaction;   by plotting  the correlation between the value of $|{\bf p}_2|$
 and the angle between ${\bf k}_1$ and
 ${\bf p}_2$ , it was found  that  whereas recoiling nucleons with momentum of the order or
 less than the Fermi momentum were emitted isotropically, nucleons with momentum  $p_2 \simeq 2-2.5\,fm^{-1}$ were emitted in a
 backward cone with respect to the direction of ${\bf k}_1$, in agreement with the picture of the
 absorption of the virtual photon by a nucleon
   with a \emph{c.m.}  distribution in $^{12}$C of the type  $n_{c.m.}(K) \propto \exp[-K_{c.m.}^2/2\sigma^2]$ with
  $\sigma=7.26\pm 0.086 \,fm^{-1}$ in  agreement with the prediction of Ref. \cite{CiofidegliAtti:1995qe}, namely  $\sigma=1/\sqrt{2\alpha}=7.1
  \,fm^{-1}$ (see Section \ref{subsec:5.3}).
  Furthermore by comparing with the same apparatus and kinematics the yield
  of $^{12}C(e,e'p)X$  with  the yield of $^{12}C(e,e'pn)X$  it has been possible to obtain
  information about the ratio of $pn$ to $pp$ correlated pairs.  A detailed discussion of these experiments and their interpretation is given in
  Ref. \cite{Arrington:2012}.
\section{Conclusions}
\label{sec:7}
{\it Ab initio} many-body calculations performed in terms of realistic bare two-nucleon
interactions show that two-nucleon short-range correlations (2N SRCs), characterized by the presence of a correlation hole in the two-nucleon density in nuclei,
 exhibit a universal character,  manifesting  itself in several A-independent
features of nucleon momentum distributions. As a matter of fact,  the calculated two-nucleon relative density displays
a correlation hole which is  essentially  independent of the mass  of the nucleus, a feature that demonstrates
that two-nucleon motion at short relative distances  is
practically unaffected by the motion of nearby nucleons. This universal behavior in coordinate space reflects itself
 in peculiar universal features  of  one-nucleon,  $n_A(|{\bf k}|)$,
 and two-nucleon, $n^{N_1N_2}(|{\bf k}_{rel}|, |{\bf K}_{c.m.}|, \Theta)$, momentum distributions.
 Concerning $n_A(|{\bf k}|)$, 2N SRCs increase the high-momentum part   by orders of magnitude with respect to
 MF predictions; as for $n^{N_1N_2}(|{\bf k}_{rel}|, |{\bf K}_{c.m.}|, \Theta)$,
 particularly worth being stressed again
is  the following main feature  characterizing the motion of a $pn$ pair in medium: in the SRCs region, where
2 $\lesssim {k}_{\text{rel}}\lesssim $5 fm$^{-1}$ and, at the same time,   ${K}_{\text{c.m.}}\lesssim$ 1
fm$^{-1}$,  the  relative  and \emph{c.m.} motions of the pair are decoupled,
with the former described by a  deuteron-like
momentum distribution,  and the latter,  governing the A dependence of the motion, described by a  momentum
distribution linked to
the average value of the MF kinetic energy. Such a decoupling of the relative and \emph{c.m.}
momenta have been theoretically justified by many-body calculations which predict
 factorization of the nuclear wave function at short inter-nucleon distances or, equivalently, at
high values of $k_{rel}$ and  low values of $K_{c.m.}$.
Some aspects of this picture have already been partially confirmed by experiments providing evidence on
the high-momentum content  of the one-nucleon momentum distribution, by the experimental behavior of the inclusive
 electron-nucleus cross section ratios and, eventually,  by the measurement of
 the percentage ratio of $pn$ to $pp$  correlated pairs in $^4$He and $^{12}$C.
Much work however remains to be done in
order to investigate the three-dimensional structure of  the two-nucleon momentum distributions
$n^{N_1N_2}(|{\bf k}_{rel}|, |{\bf
K}_{c.m.}|, \Theta)$, with particular attention to its   \emph{c.m.} dependence in the SRCs region, as well as to its structure in  the region
where both $k_{rel}$ and $K_{c.m.}$ are large, characterized by the breaking down of c.m and relative momenta factorization due to
the expected dominant role of   many-nucleon SRCs.

Unveiling the correlation structure of nuclei is a fundamental task
of nuclear physics, for by this way  information on the basic in-medium NN interaction  can be obtained.
Moreover, it should also be stressed that, recently,
a non   trivial impact of 2N SRCs on different fields, such as high-energy hadron-nucleus
\cite{totalnA,nashboris,CiofidegliAtti:2011fh} and nucleus-nucleus scattering
\cite{Alvioli:2010yk}, deep inelastic scattering \cite{Hen:2013} and  the
equation of state of unconventional nuclear matter \cite{Frick:2004th,Frankfurt:2008zv}, has been demonstrated.


\end{document}